\documentclass[aps,prl,superscriptaddress,floatfix,onecolumn, notitlepage, preprint,nobibnotes]{revtex4-2}
\usepackage{amsmath,amssymb,amsfonts}
\usepackage{hyperref}
\usepackage{amssymb}
\usepackage{graphicx}
\usepackage{ulem}
\usepackage[table,x11names]{xcolor}
\usepackage{textcomp, gensymb}
\usepackage{bm}
\usepackage{physics}
\usepackage{colortbl}
\usepackage{booktabs}
\usepackage[left]{lineno}

\makeatletter

    \def\CT@@do@color{%
      \global\let\CT@do@color\relax
            \@tempdima\wd\z@
            \advance\@tempdima\@tempdimb
            \advance\@tempdima\@tempdimc
    \advance\@tempdimb\tabcolsep
    \advance\@tempdimc\tabcolsep
    \advance\@tempdima2\tabcolsep
            \kern-\@tempdimb
            \leaders\vrule
                    \hskip\@tempdima\@plus  1fill
            \kern-\@tempdimc
            \hskip-\wd\z@ \@plus -1fill }
    \makeatother

\makeatletter
\def\blfootnote{\xdef\@thefnmark{}\@footnotetext}
\makeatother

\makeatletter
\newcommand{\fmarki}{*}
\newcommand{\fmarkii}{\ensuremath{\dagger}}
\newcommand{\fmarkiii}{\ensuremath{\ddagger}}
                
\def\@fnsymbol#1{{\ifcase#1\or \fmarki\or \fmarkii\or \fmarkiii \else\@ctrerr\fi}}
\makeatother
\renewcommand{\fmarki}{\ensuremath{\dagger}}
\renewcommand{\fmarkii}{\ensuremath{\ddagger}}
\renewcommand{\fmarkiii}{\ensuremath{\mathsection}}

\begin{document}


\title{Strain-tunability of the multipolar Berry curvature in altermagnet MnTe}

\author{Shane Smolenski}
\affiliation{Department of Physics, University of Michigan, Ann Arbor, MI 48109, USA}

\author{Ning Mao}
\affiliation{Max Planck Institute for Chemical Physics of Solids, 01187 Dresden, Germany}

\author{Dechen Zhang}
\affiliation{Department of Physics, University of Michigan, Ann Arbor, MI 48109, USA}

\author{Yucheng Guo}
\affiliation{Department of Physics and Astronomy, Rice University, Houston, TX 77005, USA}

\author{A.K.M. Ashiquzzaman Shawon}
\affiliation{Department of Physics, University of Michigan, Ann Arbor, MI 48109, USA}

\author{Mingyu Xu}
\affiliation{Department of Chemistry, Michigan State University, East Lansing, MI 48864, USA}

\author{Eoghan Downey}
\affiliation{Department of Physics, University of Michigan, Ann Arbor, MI 48109, USA}

\author{Trisha Musall}
\affiliation{Department of Physics, University of Michigan, Ann Arbor, MI 48109, USA}

\author{Ming Yi}
\affiliation{Department of Physics and Astronomy, Rice University, Houston, TX 77005, USA}

\author{Weiwei Xie}
\affiliation{Department of Chemistry, Michigan State University, East Lansing, MI 48864, USA}

\author{Chris Jozwiak}
\affiliation{Advanced Light Source, Lawrence Berkeley National Laboratory, Berkeley, CA 94720, USA}

\author{Aaron Bostwick}
\affiliation{Advanced Light Source, Lawrence Berkeley National Laboratory, Berkeley, CA 94720, USA}

\author{Nobumichi Tamura}
\affiliation{Advanced Light Source, Lawrence Berkeley National Laboratory, Berkeley, CA 94720, USA}

\author{Eli Rotenberg}
\affiliation{Advanced Light Source, Lawrence Berkeley National Laboratory, Berkeley, CA 94720, USA}

\author{Lu Li}
\affiliation{Department of Physics, University of Michigan, Ann Arbor, MI 48109, USA}

\author{Kai Sun}
\affiliation{Department of Physics, University of Michigan, Ann Arbor, MI 48109, USA}

\author{Yang Zhang}
\affiliation{Department of Physics and Astronomy, University of Tennessee, Knoxville, TN 37996, USA}
\affiliation{Min H. Kao Department of Electrical Engineering and Computer Science, University of Tennessee, Knoxville, Tennessee 37996, USA}

\author{Na Hyun Jo}
\altaffiliation{nhjo@umich.edu}
\affiliation{Department of Physics, University of Michigan, Ann Arbor, MI 48109, USA}

\date{\today}

\def\kill #1{\sout{#1}}
\def\add #1{\textcolor{blue}{#1}} 
\def\addred #1{\textcolor{red}{#1}} 

\maketitle

\section{Abstract}
The anomalous Hall effect describes the generation of a transverse voltage by a longitudinal current even in the absence of an external magnetic field. While typically observed in ferromagnets, it has also been predicted to arise in altermagnets, materials characterized by rotational symmetries that enable broken time reversal symmetry despite compensated collinear magnetic ordering. These symmetries enforce band (anti)crossings that can generate significant contributions to the Berry curvature that drives the anomalous Hall effect. This Berry curvature is predicted to exhibit a characteristic multipolar order, resulting in a symmetry-enforced distribution at or near net compensation which is highly sensitive to perturbations that distort this balance. However, exploring the predicted multipolar Berry curvature of altermagnets and its reversible manipulation remains challenging. Here, we demonstrate evidence for the multipolar nature of the altermagnetic Berry curvature in MnTe by tuning the anomalous Hall effect via uniaxial stress. Upon straining, the magnitude of the anomalous Hall conductivity changes and, at a critical strain of 0.14\%, the sign is reversed. Symmetry analysis and density functional theory calculations reveal that this tunability is a direct consequence of the altermagnetic multipolar Berry curvature. Our results provide insight into the role of crystal and magnetic symmetries in the realization of higher-order Berry curvature distributions and their unique tunability. 


\section{main}
The Hall current $\textbf{j}_H$, the transverse current in solids arising from a longitudinal electric field $\textbf{E}$, can be described as $\textbf{j}_H = \textbf{h} \times \textbf{E}$\add{,} where $\textbf{h}=(\sigma_{zy},\sigma_{xz},\sigma_{yx})$ is the Hall vector consisting of the antisymmetric Hall conductivity components $\sigma_{ij}$. Because $\textbf{j}_H$ is odd under time reversal while $\textbf{E}$ is even, $\textbf{h}$ is odd under time reversal. Therefore, $\textbf{h}$, and thus $\textbf{j}_H$, can only arise in systems where time reversal symmetry is broken. For the ordinary Hall effect, an external magnetic field is the source of the time reversal symmetry breaking and the resulting Lorentz force on charge carriers generates the transverse current. For the anomalous Hall effect (AHE), however, the crystal symmetries themselves must break time reversal symmetry. A transverse current can then arise from extrinsic scattering processes or an intrinsic mechanism driven by non-trivial net Berry curvature throughout the Brillouin zone~\cite{NAGAOSA2010}. Conventionally, the net magnetization of ferromagnets (FMs) drives the broken time reversal symmetry for the AHE. Although recent experiments have demonstrated the AHE in other systems, including non-collinear magnets~\cite{NAKATSUJI2015, CHEN2014, NAYAK2016,zhang2017strong,MACHIDA2010} and topological materials~\cite{CHANG2023}, it has long been assumed that, in collinear magnetic systems, net magnetization is necessary for the bulk AHE.

However, a third collinear magnetic order alongside FM and antiferromagnetism, known as altermagnetism (AM), breaks time reversal symmetry through crystal symmetries alone without net magnetization or spin-orbit coupling (SOC)~\cite{ SMEJKAL2022_1, SMEJKAL2022_2, SMEJKAL2022_3}. More specifically, AMs are characterized by two opposite spin sublattices connected by either proper or improper rotational symmetries. These symmetries lead to spin-up and spin-down bands that are generally non-degenerate outside of symmetry-enforced nodal planes and form an alternating order (d-wave or higher even parity wave). These symmetry-enforced band (anti)crossings occurring at the nodal planes can generate significant contributions to the Berry curvature, which, as a result, is also predicted to generally exhibit multipolar order with an $s$-wave component that can be zero or non-zero depending on the crystal symmetries~\cite{SMEJKAL2022_1,SMEJKAL2022_2,SMEJKAL2022_3,TAKAHASHI2025}. This differs from ferromagnets, where Berry curvature hotspots arising from band (anti)crossings generally rely on microscopic details of the electronic structure~\cite{SMEJKAL2022_3}. Additionally, the presence of this multipole implies that, at or near compensation, small perturbations can distort the Berry curvature distribution. This, in turn,  generates large non-zero integrated values, and pronounced changes to the AHE, with the sign dependent on the direction of the perturbation~\cite{TAKAHASHI2025}. This presents a means of reversibly tuning the Berry curvature and AHE without significantly altering the magnetic order, in contrast to other AHE-tunable FM and non-collinear magnetic systems~\cite{IKHLAS2022,TIAN2021,SAMATHRAKIS2020,ZHOU2025}. Furthermore, since the AHE is accompanied by weak ferromagnetism (WF) in altermagnets, careful tuning of the AHE that can manipulate the Berry curvature while minimally affecting the magnetic order can help in disentangling the intrinsic altermagnetic origin of the AHE in AMs. Additionally, tunability of the electronic properties of AMs - which feature vanishing net magnetization, collinear order, and terahertz switching dynamics - is essential for development into next-generation spintronic applications~\cite{SMEJKAL2022_1, JUNGWIRTH2016,JUNGWIRTH2025}. Yet, detecting signatures of this multipolar Berry curvature and realizing its reversible, magnetic field-free control remains challenging.

Since crystal symmetry is at the heart of AM, uniaxial stress provides a powerful mechanism to tune AMs and control the AHE. Strain generated by a uniaxial stress can lower the rotational symmetry of the crystal, affecting the AM splitting in the electronic structure and the associated altermagnetic properties~\cite{BELASCHENKO2025,KARETTA2025, OGAWA2025,PENG2025}. Importantly, it has also been predicted that the AHE in AMs can directly be tuned by strain through the multipolar Berry curvature order, a phenomenon referred to as the piezo-Hall effect~\cite{TAKAHASHI2025}. In this study, we manipulate both the magnitude and sign of the AHE in altermagnet MnTe using uniaxial stress without changing the magnetization direction.
Through density functional theory (DFT) calculations and symmetry arguments, we demonstrate that the observed tunability can be ascribed to a direct tuning of the multipolar Berry curvature. Our results demonstrate that the AHE in AMs is highly sensitive to external perturbations and rooted in the non-trivial Berry curvature multipole. More broadly, this finding further supports the intrinsic origin of time reversal symmetry breaking phenomena observed in AM candidates.

\section{Crystal symmetry and AHE in MnTe}
\begin{figure} [ht]
    \includegraphics[width = \textwidth]{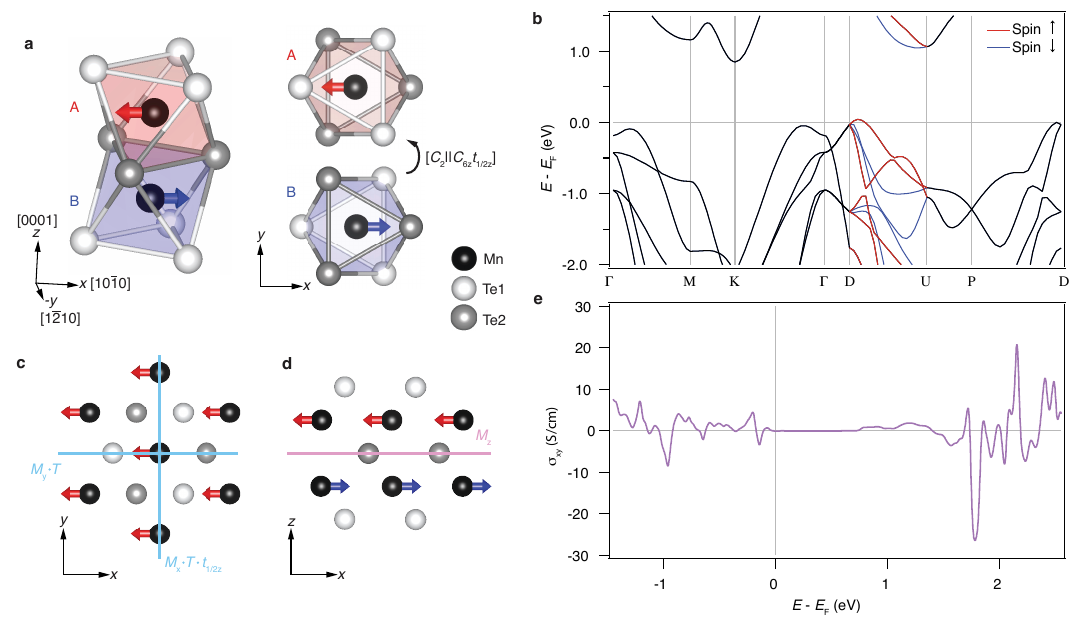}%
    \caption{\textbf{Time-reversal symmetry breaking and Hall vector in MnTe.} \textbf{a}, Crystal structure of MnTe. A six-fold rotation about $[0001]$ (\textit{z}-axis)  coupled with a half unit cell translation about the \textit{z}-axis in real space along with a reversal of the spins connects the up/down spin sublattices, breaking time-reversal symmetry. \textbf{b}, Electronic band structure of MnTe as calculated by DFT without spin-orbit coupling. Red/blue lines denote spin up/down bands while black lines denote spin-degenerate bands. In the $k_z=\frac{\pi}{2c}$ plane (DUPD plane; See Supplementary Note 1 for labeled Brillouin zone), the spin up/down bands can be non-degenerate. \textbf{c},\textbf{d}, Crystal mirror $M$ and magnetic mirror $M\cdot T$ symmetries of MnTe. As a result, the Hall vector is allowed parallel to the \textit{z}-axis. \textbf{e}, Calculated anomalous Hall conductivity $\sigma_{xy}$ as a function of Fermi level position in MnTe.
    \label{fig:Structure}}
\end{figure}

$\alpha$-MnTe (hereon MnTe) is a prominent AM material candidate that crystallizes in the $P6_3/mmc$ space group (number 194) (Fig.~\ref{fig:Structure}a). Upon cooling below the Néel temperature of 307\,K~\cite{KLUCZYK2024,SQUIRE1939,UCHIDA1956} (Supplementary Note 1), the Mn magnetic moments form an \textit{A}-type antiparallel magnetic ordering with the easy axis along $<\!10\bar10\!>$ (magnetic point group m’m’m, number 8.4.27)~\cite{KRIEGNER2017, KOMATSUBARA1963}. In the magnetically ordered phase, the two sublattices are connected by a 6-fold crystal rotation about [0001] (\textit{z} -axis) coupled with a half-unit cell translation along the \textit{z}-axis and a 2-fold spin rotation (Fig.~\ref{fig:Structure}a). This is represented by the symmetry operator $[C_2||C_{6z}t_{\frac{1}{2}z}]$, where the operator on the left of the double bars acts on spin space while the operators on the right act on real space. The rotational symmetry connecting the two sublattices renders MnTe as an AM, breaking time reversal symmetry. Accordingly, DFT calculations of the electronic band structure of MnTe predict a large, alternating spin-splitting along low-symmetry paths within the Brillouin zone (Fig.~\ref{fig:Structure}b). The spin splitting of the bands is one of the largest among predicted AMs~\cite{SMEJKAL2022_2} and recent angle-resolved photoemission spectroscopy studies have found evidence of such splitting~\cite{LEE2024,KREMPASKY2024}. Furthermore, evidence for time-reversal symmetry breaking responses in MnTe have also been demonstrated with inelastic neutron scattering~\cite{LIU2024} and X-ray dichroism experiments~\cite{AMIN2024, HARIKI2024, TAKEGAMI2025,YAMAMOTO2025}, although these signatures have not been definitively distinguished from relativistic effects such as SOC or WF.

Beyond requiring time reversal symmetry breaking, the emergence of the AHE in altermagnets also depends on the direction of the Néel vector $\textbf{L}$ and the associated mirror symmetries~\cite{ZHOU2025}. Notably, the symmetries of MnTe for $\textbf{L}\parallel [10\bar10]$ enable a non-zero $\textbf{h}$.
Specifically, $(10\bar10)$ and $(1\bar210)$ are magnetic mirrors, enforcing the axial vector $\textbf{h}$ parallel to these planes (Fig.~\ref{fig:Structure}c), while (0001) is a crystal mirror, requiring $\textbf{h}$ to be perpendicular to this mirror plane (Fig.~\ref{fig:Structure}d). Thus, $\textbf{h}\parallel [0001]$ is allowed by symmetry. As a result, the anomalous Hall conductivity in the \textit{xy}-plane can be non-zero, as demonstrated via our DFT calculations in Fig.~\ref{fig:Structure}e. Importantly, with $\textbf{h}\parallel z$, the Hall current is perpendicular to $z$ and is thus experimentally measureable. WF is also allowed along the \textit{z}-axis by the same symmetries that enable a non-zero $\textbf{h} \parallel z$. While this complicates the origin of the AHE in MnTe, this net magnetization also provides a mechanism to flip the Néel vector, and the AHE, with an applied magnetic field~\cite{KLUCZYK2024}.

To explore the AHE in MnTe, we first performed Hall measurements on a single-crystalline sample (Sample A)  (Ambient mounting condition) at various temperatures (Methods). After subtraction of the ordinary Hall response and symmetrization (Supplementary Note 2), a clear non-zero, hysteretic anomalous Hall resistivity $\rho_{xy}^{\text{AHE}}$ is observed below $T_N$ (Fig.~\ref{fig:Thermal_Strain}c). This hysteretic behavior with magnetic field $H$ corresponds to switching between domains with opposite $\bf{L}$ and, correspondingly, opposite $M_z$~\cite{KLUCZYK2024,AMIN2024}. As temperature is further lowered, the hysteretic behavior persists without changing sign although the magnitude of the saturated $\rho_{xy}^{\text{AHE}}$ gradually diminishes. These results agree with previous measurements performed on single crystals of MnTe~\cite{KLUCZYK2024} although the hysteresis of $\rho_{xy}^{\text{AHE}}$  has the opposite sign as that measured in epitaxially-strained thin films~\cite{GONZALEZ2023}. This observation, along with previously reported large magnetostriction~\cite{BARAL2023}, piezomagnetism~\cite{AOYAMA2024}, and predictions of tunable electronic structure and spin current generation under stress~\cite{BELASCHENKO2025,DEVARAJ2024,CHEN2025,ALAEI2025, CARLISLE2025}, suggest strong coupling between lattice, electronic, and magnetic degrees of freedom in MnTe, rendering MnTe as a promising candidate for exploring the piezo-Hall effect.

\section{Tuning of the AHE via Uniaxial Stress}
\begin{figure} [ht]
    \includegraphics[width = \textwidth]{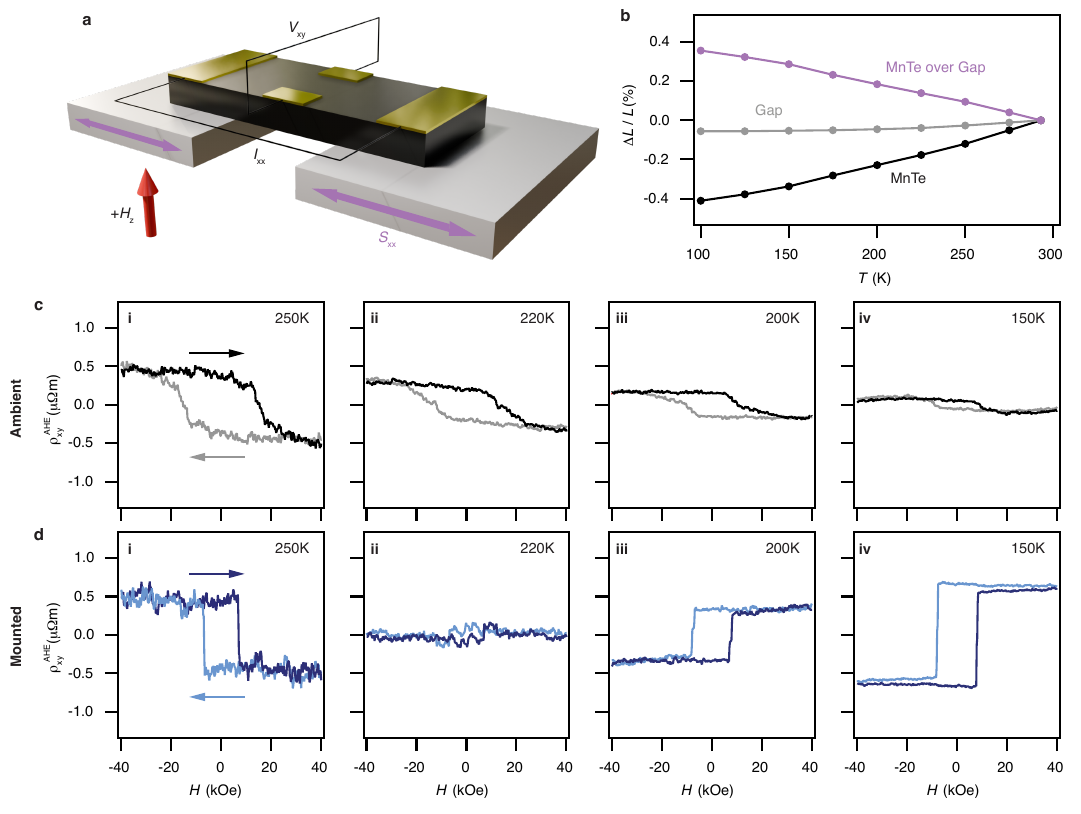}%
    \caption{\textbf{Switching of the AHE through differential thermal contraction.} \textbf{a}, Schematic of the experimental setup to measure strain-dependence of the AHE. An MnTe sample with sputtered Au contact pads is mounted across two movable sample plates of a uniaxial stress cell. The current $I_{xx}$ and stress $S_{xx}$ are applied along the \textit{x}-axis ($[10\bar10]$), the Hall voltage $V_{xy}$ is measured along the \textit{y}-axis, and an external magnetic field $H_z$ is applied along the \textit{z}-axis. Strain is generated both mechanically (by moving the sample plates) and thermally (through differential thermal contraction between sample and sample plates). \textbf{b}, Relative change in length as a function of temperature for MnTe~\cite{BARAL2023} (black curve), the gap between the sample plates on our stress cell (gray curve), and an MnTe sample mounted across the gap of our stress cell (purple curve). \textbf{c},\textbf{d}, Anomalous Hall resistivity $\rho_{xy}^{\text{AHE}}$ as a function of external magnetic field $H$ at various temperatures for the Ambient (\textbf{c}) and Mounted (\textbf{d}) conditions for Sample A. The direction of the field sweeps are indicated by the arrows in panel \textbf{i} of (\textbf{c}) and (\textbf{d}).
    \label{fig:Thermal_Strain}}
\end{figure}

To probe the strain-dependence of the AHE in MnTe, we employed a mechanically-driven Ti stress cell (Razorbill Instruments MC050). By mounting a sample across a gap between two movable plates on the cell (Fig.~\ref{fig:Thermal_Strain}a), strain can be generated via two distinct mechanisms. First, mechanical strain $\varepsilon^{M}$ is induced by mechanically moving the sample plates and thus varying the gap/sample length. Additionally, thermal strain $\varepsilon^{T}$ develops as temperature is varied due to a difference between the coefficients of thermal expansion for MnTe and the stress cell itself (Fig.~\ref{fig:Thermal_Strain}b). For MnTe, which has a relatively large coefficient of thermal expansion~\cite{BARAL2023} compared with that of Ti~\cite{HIDNERT1943}, thermal tensile strain gradually develops with decreasing temperature, reaching a value of $\varepsilon_{xx}^T\approx0.35\%$ at 100\,K. The strains generated by the cell are calibrated via a strain gauge and effective transmission to an MnTe sample is confirmed via X-ray microdiffraction (Supplementary Note 3). Each of $\varepsilon^{T}$ and $\varepsilon^{M}$ can be varied independently and their sum represents the total strain $\varepsilon$ experienced by the sample.

We first explored the effects of thermal strain by measuring $\rho_{xy}^{\text{AHE}}$ at various temperatures for a sample mounted on the stress cell with $\varepsilon^M_{xx}=0.00\%$ (Mounted condition) (Fig.~\ref{fig:Thermal_Strain}d). Sample A was mounted across the gap of the cell so as to apply stress $\mathbf{S}$ along $[10\bar10]$ (\textit{x}-direction), $\mathbf{S} \parallel \hat x$ (Fig.~\ref{fig:Thermal_Strain}a). At 250\,K, the behavior of $\rho_{xy}^{\text{AHE}}$ is in good agreement with that measured in the Ambient condition in both magnitude and sign. The transition between the two $\textbf{L}$ states is sharper for the Mounted condition than the Ambient condition, possibly due to domain alignment arising from thermal strain. Interestingly, upon lowering the temperature to 220\,K, $\rho_{xy}^{\text{AHE}}$ is nearly entirely quenched. As temperature is further lowered ($\varepsilon^{T}_{xx}$ is increased), the hysteretic behavior of $\rho_{xy}^{\text{AHE}}$ reemerges but with a reversed sign and a magnitude that increases monotonically with decreasing temperature, a distinct departure from Ambient condition measurements. As the only difference between the Ambient and Mounted measurements is the mounting condition, $\varepsilon^{T}$ is likely the origin of the sign reversal of the AHE.

\begin{figure} [ht]
    \includegraphics[width = 0.5\textwidth]{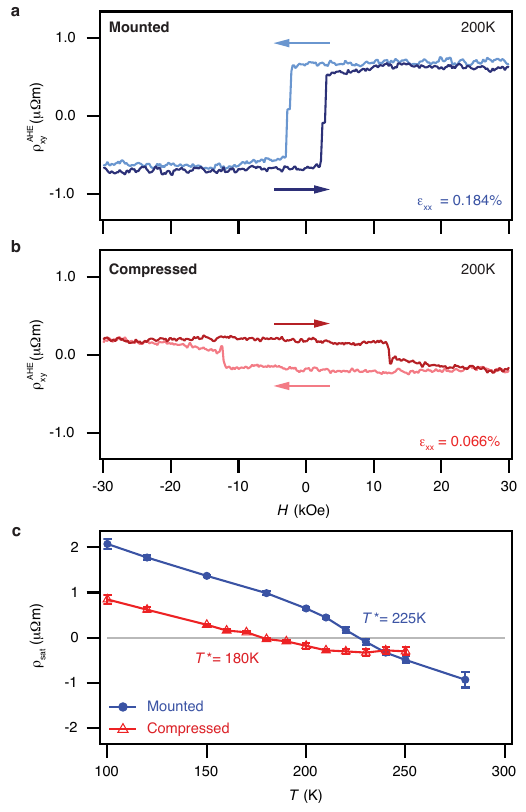}%
    \caption{\textbf{Applied mechanical strain tunability of the AHE.} \textbf{a},\textbf{b}, Anomalous Hall resistivity $\rho_{xy}^{\text{AHE}}$ as a function of external magnetic field $H$ at 200K in the Mounted (\textbf{a}) and Compressed (\textbf{b}) conditions for Sample B. The total strain ($\varepsilon_{xx}=\varepsilon_{xx}^T+\varepsilon_{xx}^M$) experienced in each condition is labeled in the plots and arrows indicate direction of field sweeps. \textbf{c}, Averaged saturated anomalous Hall resistivity $\rho_{sat}$ in the $+H$ polarized state as a function of temperature for the Mounted (blue curve) and Compressed (red curve) conditions with the transition temperature $T^*$ labeled for each.
    \label{fig:Mechanical_Strain}}
\end{figure}

To isolate the effects of strain, we subsequently measured the AHE under compressive mechanical stress with $\varepsilon^{M}_{xx}=-0.118\%$ (Compressed condition). Figures.~\ref{fig:Mechanical_Strain}a,b show $\rho_{xy}^{\text{AHE}}$ measured at 200\,K for two different values of applied strain on Sample B (with the same orientation/setup as Sample A). For the Mounted condition ($\varepsilon^{M}_{xx}=0.00\%$, $\varepsilon^{T}_{xx}=0.184\%$, total strain $\varepsilon_{xx}=0.184\%$), the sign of $\rho_{xy}^{\text{AHE}}$ is positive at large positive $H$ and negative for large negative $H$ (Fig.~\ref{fig:Mechanical_Strain}a). For the Compressed condition ($\varepsilon_{xx}=0.066\%$), on the other hand, the hysteresis is reversed, matching the sign observed under the Ambient condition at 200\,K (Fig.~\ref{fig:Mechanical_Strain}b). Therefore, at a constant temperature, uniaxial tensile stress is capable of switching the sign of the hysteretic behavior of the AHE alone. 

We also compare the saturated $\rho_{xy}^{\text{AHE}}$ at positive magnetic fields $\rho_{sat}$ for the Mounted and Compressed condition as a function of temperature (Fig.~\ref{fig:Mechanical_Strain}c). For both conditions, decreasing temperature (increasing $\varepsilon^T$) and increasing $\varepsilon^M$ shifts $\rho_{\text{sat}}$ towards more positive values for all temperatures below 240\,K, demonstrating the significant impact of strain on the AHE over a broad region. Additionally, we observe that the critical temperature $T^*$ of the hysteresis reversal shifts from $T^*\approx 225$\,K for the Mounted condition to $T^*\approx 180$\,K for the Compressed condition while the overall trends remain similar. This behavior can be understood as the compressive mechanical strain partially cancelling the tensile thermal strain that develops with decreasing temperature. Moreover, upon release of the mechanical strain, the temperature dependence of $\rho_{xy}^{\text{AHE}}$ reverts back to that of the Mounted condition (Supplementary Note 4). These observations further establish the underlying role of strain as the driving factor in the hysteresis reversal.

\section{Electronic Origin of the Reversible AHE}
The strain-induced changes to both the sign and magnitude of the AHE in MnTe can likely be ascribed to direct tuning of the altermagnetic Berry curvature multipole. To clarify this, we consider that the anomalous Hall conductivity $\sigma_{xy}$ in AM with WF can be separated into two components, one arising from WF and the other from AM order. We thus write $\sigma_{xy}$ as a function of both the AM order parameter (the Néel vector $\bf{L}$) and the WF order parameter (the net magnetization $M_z$), $\sigma_{xy}(\textbf{L},M_z)$. In the ground state, the sign of $M_z$ is locked to the orientation of $\bf{L}$, but here we consider the more general case in which they can vary independently. We can then decompose $\sigma_{xy}(\textbf{L},M_z)$ into two symmetry-distinct contributions
\begin{equation}
    \sigma_{xy}=\sigma_{xy}^{\text{AM}}+\sigma_{xy}^{\text{WF}},
\end{equation}
where 
\begin{equation}
    \sigma_{xy}^{\text{AM}}(\textbf{L},M_z)=\frac{\sigma_{xy}(\textbf{L},M_z)+ \sigma_{xy}(\textbf{L},-M_z)}{2}
\end{equation}
and
\begin{equation}
    \sigma_{xy}^{\text{WF}}(\textbf{L},M_z)=\frac{\sigma_{xy}(\textbf{L},M_z)- \sigma_{xy}(\textbf{L},-M_z)}{2}.
\end{equation}
$\sigma_{xy}^{\text{AM}}$ is odd in $\bf{L}$ (even in $M_z$) and represents the contribution to $\sigma_{xy}$ arising from the AM order while $\sigma_{xy}^{\text{WF}}$ is odd in $M_z$ (even in $\bf{L}$) and represents the contribution to $\sigma_{xy}$ arising from WF~\cite{SMEJKAL2020}. Importantly, both $\sigma_{xy}^{\text{WF}}$ and $\sigma_{xy}^{\text{AM}}$ are functions of strain with different microscopic origins.

To understand the microscopic mechanism driving the strain-dependent AHE hysteresis, we consider two effects. First, strain may directly couple to $\sigma_{xy}^{\text{AM}}$ by modifying the electronic structure, changing the Berry curvature compensation and thus altering its integrated magnitude and sign~\cite{TAKAHASHI2025}. This strain-induced reversal of $\sigma_{xy}^{\text{AM}}$ without changing the magnetic ordering describes the piezo-Hall effect, which can be expressed as
\begin{equation}
    \sigma_{\alpha\beta}^{\text{AM}}(\varepsilon) = \sigma_{\alpha\beta}^{\text{AM}}(\varepsilon=0) + \nu_{\alpha\beta\gamma\delta}\varepsilon_{\gamma\delta},
    \label{eq:elastoHall}
\end{equation}
where $\nu_{\alpha\beta\gamma\delta}$ is the piezo-conductivity tensor. Notably, the magnetic point group of MnTe allows for $\nu_{xyxx}$, $\nu_{xyyy}$, and $\nu_{xyzz}$ to be non-zero, demonstrating that $\mathbf{S} \parallel \hat x$, which generates $\varepsilon_{xx}$, $\varepsilon_{yy}$, and $\varepsilon_{zz}$, may indeed alter $\sigma_{xy}$ in MnTe~\cite{TAKAHASHI2025}. For a large enough $\nu_{\alpha\beta\gamma\delta}$, the piezo-Hall effect can then account for changes to both the magnitude and sign of the AHE, as seen in our experiments.

Alternatively, the hysteresis reversal may be generated by strain-dependent changes to the magnetic ordering via piezomagnetism, which, in turn, can affect $\sigma_{xy}^{\text{WF}}$ and $M_z$. Piezomagnetism can be described as
\begin{equation}
    M_\mu(\varepsilon)=M_\mu(\varepsilon=0)+\Lambda_{\mu\gamma\delta}\varepsilon_{\gamma\delta},
\end{equation}
where $\Lambda_{\mu\gamma\delta}$ is the piezomagnetic tensor. The same symmetries that enabled non-zero terms of the piezo-conductivity tensor also enable non-zero $\Lambda_{zxx}$, $\Lambda_{zyy}$, and $\Lambda_{zzz}$, demonstrating that $\mathbf{S} \parallel \hat x$ can control $M_z$ in MnTe. Either a sign reversal of $M_z$ that does not flip the sign of $\sigma_{xy}$ or modifications to the electronic structure arising from the changed $M_z$ and which tune $\sigma_{xy}^{\text{WF}}$ could, in principle, describe the observed strain-dependent hysteresis reversal.

However, we argue that piezomagnetism cannot explain both the changes in sign and magnitude of the AHE observed in experiment. First, the exceedingly small value of $M_z$ (Supplementary Notes 1 and 5) suggests that $\sigma_{xy}^{\text{WF}} \approx 0$ and is thus unlikely to account for the sizeable AHE~\cite{KLUCZYK2024,GONZALEZ2023}. Furthermore, based on previous measurements of the piezomagnetic tensor~\cite{AOYAMA2024}, $M_z$ cannot be considerably enhanced for the levels of stress achieved in our study, making it unlikely for $\sigma_{xy}^{\text{WF}}$, and $\sigma_{xy}^{\text{AM}}$, to be significantly altered by strain-induced magnetization. On the other hand, the small value of $M_z$ makes it possible for its sign to be reversed under small values of strain. However, while a sign reversal of $M_z$ (that does not change the sign of $\sigma_{xy}$) may describe the change in sign of the hysteresis, that alone cannot account for the strain-induced changes in magnitude of the AHE; since $\bf{h}$ is fixed along the $z$-axis and $\sigma_{xy}^{\text{WF}} \approx 0$, changes to the magnitude of $\sigma_{xy}$ under strain must be derived from $\sigma_{xy}^{\text{AM}}$, already demonstrating the significant strain-tunability of $\sigma_{xy}^{\text{AM}}$. Furthermore, we find that strain is capable of switching the sign of $\sigma_{xy}$ at zero magnetic field, confirming that a sign change of $\sigma_{xy}^{\text{AM}}$, and not $M_z$, is responsible for the observed hysteresis reversal in the AHE (Supplementary Note 6). Therefore, while piezomagnetism is symmetry-allowed, it cannot describe the observed strain-induced changes to the AHE.

\begin{figure} [ht]
    \includegraphics[width = \textwidth]{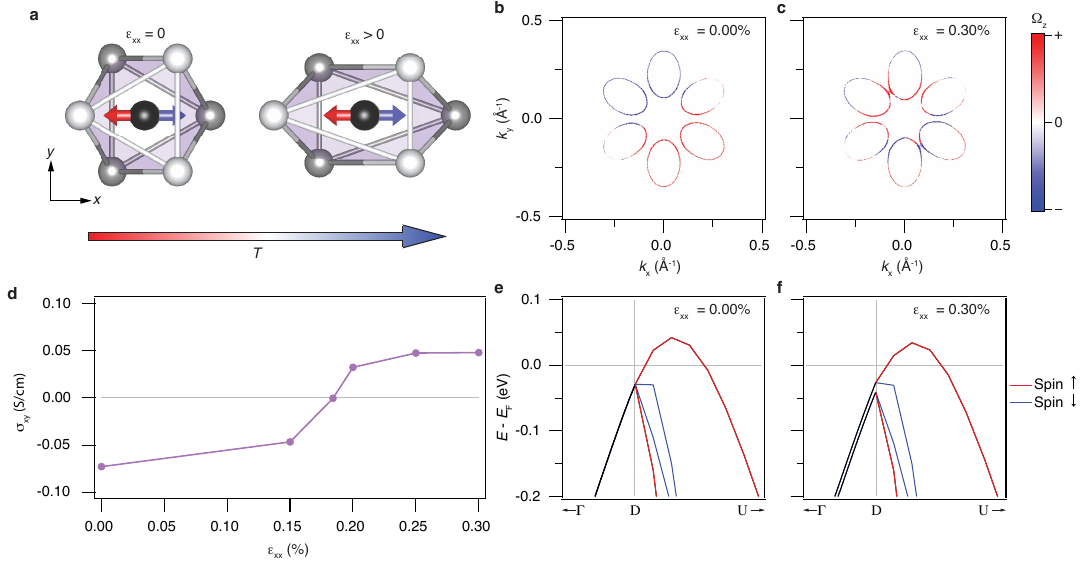}%
    \caption{\textbf{Electronic origin of the tunable AHE.} \textbf{a}, Schematic illustrating the effects of uniaxial stress on the unit cell of MnTe. Tensile strain is experienced along the \textit{x}-axis while compressive strain is experienced along the \textit{y}- and \textit{z}-axes. The arrow depicts how tensile stress along the $x$-direction develops as temperature decreases for samples mounted on the stress cell. \textbf{b},\textbf{c}, Calculated Berry curvature $\Omega_z$ in momentum space as a function of ($k_x$,$k_y$) for unstrained (\textbf{b}) and strained (\textbf{c}) MnTe. The sign of the integrated Berry curvature is opposite for the two cases. \textbf{d}, Calculated anomalous Hall conductivity $\sigma_{xy}$ as a function of uniaxial stress. \textbf{e},\textbf{f}, Calculated electronic structure near D along the $\Gamma$-D-U path for unstrained (\textbf{e}) and strained (\textbf{f}) MnTe. 
    \label{fig:DFT}}
\end{figure}

To explore whether changes to the altermagnetic Berry curvature, driven by the piezo-Hall effect, can adequately describe the strain-dependence of the AHE, we calculated $\sigma_{xy}$ as a function of stress. The magnetic moments are fixed along $[10 \bar 10]$ (describing no WF and a single domain) so as to isolate $\sigma_{xy}^{\text{AM}}$, and the Fermi level was set to match the value determined by angle-resolved photoemission spectroscopy (Supplementary Note 1). The stress dependence is investigated by increasing the unit cell length along the \textit{x}-direction by a fixed amount and then reducing the lengths along the \textit{y}- and \textit{z}-directions by the corresponding amounts as calculated using the theoretical elastic tensor (Fig.~\ref{fig:DFT}a and Supplementary Note 3). Notably, the sign of $\sigma_{xy}^{\text{AM}}$ switches at a critical strain $\varepsilon_{xx}^* = 0.184\%$, which is within the error bars of the experimental value of $\varepsilon_{xx}^* = 0.14\%$ (Fig.~\ref{fig:DFT}d). As $\textbf{L}$ and $M_z$ are not varied in the calculation, this thus presents an electronic, rather than magnetic, mechanism for the observed AHE switching behavior. Note that this is distinct from other examples of strain-tunable AHE where changed magnetic ordering is responsible for and/or accompanies the tunability~\cite{IKHLAS2022,TIAN2021,SAMATHRAKIS2020,ZHOU2025}.

We then examine the microscopic changes to the electronic structure resulting from the applied stress. The Berry curvatures for unstrained ($\varepsilon_{xx}=0.00\%$) and strained ($\varepsilon_{xx}=0.30\% > \varepsilon_{xx}^*$) MnTe are presented in Figs.~\ref{fig:DFT}b,c, respectively. The multipolar nature of the Berry curvature distribution is evident in both conditions. However, in the unstrained case, the integration over the Brillouin zone is positive while the distribution has been distorted in the strained case, causing the integral to have opposite sign. The application of stress breaks the 6-fold rotational crystal symmetry and, as a result, lifts band degeneracies at high symmetry momenta, including near the Fermi level at the $D$ point (Figs.~\ref{fig:DFT}e,f). These distortions and lifted degeneracies can alter the Berry curvature distribution and, as the net Berry curvature in the unstrained case is already nearly compensated, can cause the reversal of the sign of the net Berry curvature. Therefore, strain is capable of directly tuning the altermagnetic multipolar Berry curvature multipole, explaining the observed reversal of the AHE in MnTe. Moreover, this piezo-Hall mechanism is independent of the WF, demonstrating the key role of the altermagnetic time reversal symmetry breaking in the anomalous Hall response and its tunability in MnTe.

\section{Conclusions and Outlook}
Our work demonstrating the tunability of both the magnitude and sign of the AHE in MnTe clearly establishes the strain-sensitivity of the Berry curvature in an altermagnet. This realization of the piezo-Hall mechanism supports the existence of the sensitive multipolar Berry curvature of altermagnets and corroborates the intrinsic time reversal symmetry breaking as the origin of the anomalous Hall effect in altermagnets. Moreover, the strain tunability reveals a magnetic field-free mechanism to control the AHE in altermagnets and which does not reverse the magnetic ordering. This not only demonstrates a means of experimentally verifying new altermagnetic candidates but also offers promise for spintronic/straintronic applications, such as magnetic field-free switches~\cite{BUKHARAEV2018,DUAN2025,PENG2025}, 4-state memory devices~\cite{BUKHARAEV2018, SHI2008}, and sensor technologies. More broadly, our work demonstrates the powerful role of strain in the tuning of altermagnets and can motivate the experimental study of other strain-induced phenomena in altermagnets such as valley polarization~\cite{LI2024}, spin current generation~\cite{BELASCHENKO2025}, and tunable magnetic phase transitions~\cite{CHAKRABORTY2024}.

\newpage
\section{Methods}
\subsection{Crystal growth and characterization}
Single crystals of MnTe were grown using the solution (self flux) growth method out of excess Te. An initial concentration, Mn\textsubscript{0.30}Te\textsubscript{0.70}, of elemental Mn (99.95\%, ThermoScientific) and Te (99.999+\%, ThermoScientific) was placed into a fritted alumina crucible~\cite{CANFIELD2016,SLADE2022} and sealed in a quartz ampoule under partial pressure of argon. The ampoule was then heated to 1000\degree C over 10 hours, held there for 10 hours, cooled down to 850\degree C over 5 hours, and then cooled to 760-775\degree C at a rate of 1.5\degree C per hour. At this temperature, the excess solution was separated from the MnTe crystals using a centrifuge. The crystal structure is confirmed by X-ray diffraction (XRD) measurement (Supplementary Note 1) and the crystallographic orientation was determined by single-crystal XRD and diffraction contrast tomography (DCT) (Supplementary Note 7). Ambient condition magnetic and electrical measurements are in agreement with previous reports, further confirming the growth of the correct phase (Supplementary Note 1).

\subsection{Stress cell and calibration}
A Razorbill MC050: Ultraminiature UHV Strain Cell was used for the strain measurements. The strain is adjusted by a mechanical screw in ambient conditions; turning this screw clockwise (counterclockwise) shifts two plates closer together (further apart), thus straining a sample mounted across the plates. The effective strain (thermal and mechanical) of the cell was measured using a strain gauge (Micro-Measurements C5K-06-S5198-350-33F) mounted across the gap of the stress cell using Loctite EA1C. Additionally, X-ray microdiffraction measurements were performed at Beamline 12.3.2 at the Advanced Light Source at Lawrence Berkeley National Laboratory to confirm that the stress applied by the cell leads to strain in our MnTe samples. Further details regarding the strain gauge and X-ray microdiffraction calibration measurements can be found in Supplementary Note 3.

\subsection{Sample preparation}
Crystals were polished into rectangular bars with cross sectional areas of approximately 200\,$\mu$m $\times$ 500\,$\mu$m with the length along the crystallographic $[10\bar10]$ direction, as determined by single crystal XRD and DCT. Sample lengths were typically around 1\,mm. Additional orientations were also measured and results can be found in Supplementary Note 8. Au was then deposited on top of the bars by sputtering deposition using an SPI-Module Sputter Coater and Al foil mask to create contact pads in a Hall bar configuration to reduce contact resistance and improve signal quality. Au wires were attached to the Au pads using Epotek H20E Ag epoxy. For ambient condition measurements, samples were affixed to Quantum Design Physical Propery Measurement System (PPMS) Dynacool Electrical Transport Option (ETO) pucks using a thin layer of GE varnish. For measurements on the stress cell, samples were mounted across the gap of the cell using Loctite EA1C. Samples on the stress cell were mounted such that the voltage contacts lay over the gap of the stress cell.

\subsection{Anomalous Hall effect measurements}
Anomalous Hall effect measurements were carried out in a Quantum Design PPMS Dynacool. Measurements were performed using the ETO with an excitation of 1\,mA and frequency of 33.57\,Hz. The current $I_{xx}$ was applied along $[10\bar10]$ (\textit{x}-direction), the magnetic field $H$ was applied along $[0001]$ (\textit{z}-direction), transverse voltage $V_{xy}$ was measured along $[1\bar210]$ (\textit{y}-direction), and stress $S_{xx}$ was generated along $[10\bar10]$ (\textit{x}-direction) (Fig.~\ref{fig:Thermal_Strain}a). For all measurements, the sample was initially cooled to 280\,K before turning the magnetic field to 60\,kOe. Then, at each measurement temperature, the field was swept from 60\,kOe to $-60$\,kOe and back to 60\,kOe while measuring the Hall resistance. The field was kept at 60\,kOe while changing temperatures. The data was then antisymmetrized between the up/down field sweeps and the ordinary (linear) Hall component was subtracted to achieve the anomalous Hall resistivity (Supplementary Note 2).

\subsection{Density functional theory calculations}
First-principles calculations were performed within the generalized-gradient approximation (GGA) using the Perdew–Burke–Ernzerhof (PBE) functional as implemented in the Vienna \textit{ab initio} simulation package (VASP)~\cite{PBE1,PBE2}. A Monkhorst–Pack $k$-point mesh of $7\times7\times5$ was adopted, the plane-wave cutoff energy was set to 500\,eV, and the total-energy convergence threshold was $10^{-6}$\,eV. To account for on-site Coulomb interactions, we employed the DFT+$U$ approach with an effective Hubbard parameter $U_{\mathrm{eff}} = 3.03$\,eV~\cite{GONZALEZ2023}. At $T=200$\,K and zero strain, the experimental lattice parameters $a = 4.138$\,\AA{} and $c = 6.679$\,\AA{} were used~\cite{BARAL2023}. The Bloch states obtained from DFT were projected onto atomic-like Wannier functions using \textsc{Wannier90}~\cite{Pizzi2020} to construct a tight-binding Hamiltonian. The anomalous Hall conductivity (AHC) was then evaluated from this Hamiltonian via the Kubo formula~\cite{vzelezny2023high}:
\begin{equation}
    \sigma_{xy}
    = -\frac{e^{2}}{\hbar}\sum_{n}
      \int_{\mathrm{BZ}}\frac{d^{3}k}{(2\pi)^{3}}\,
      f_{n}(\mathbf{k})\,
      \,\Omega^{xy}_{n}(\mathbf{k}),
\end{equation}
where $f_{n}(\mathbf{k})=\frac{1}{\exp\!\big[(E_{n\mathbf{k}}-\mu)/k_{\mathrm{B}}T\big]+1}$  is the Fermi–Dirac occupation of band n with chemical potential $\mu$ at temperature T. By sweeping $\mu$ across different energies, we obtain $\sigma_{xy}(\mu)$. $\boldsymbol{\Omega}_{n}(\mathbf{k})$ is the Berry curvature of band $n$, defined as
\begin{equation}
    \Omega^{xy}_{n}(\mathbf{k})
    = -2\,\mathrm{Im}\sum_{m\neq n}
      \frac{\langle u_{n\mathbf{k}}| \partial_{k_{x}} \hat{H} |u_{m\mathbf{k}}\rangle
            \langle u_{m\mathbf{k}}| \partial_{k_{y}} \hat{H} |u_{n\mathbf{k}}\rangle}
           {\big(E_{m\mathbf{k}}-E_{n\mathbf{k}}\big)^{2}+\Gamma^2},
\end{equation}
with $|u_{n\mathbf{k}}\rangle$ being the cell-periodic Bloch states and $\Gamma$ being the broadening parameter. The Brillouin-zone integrals were carried out on a dense $k$-mesh of $201\times201\times201$ to ensure numerical convergence.

\clearpage
\section{References}
\normalem
%

\clearpage
\section{Data availability}
Data underlying these results will be made available upon publication. 

\section{Acknowledgements}

N.H.J. acknowledges support from the National Science Foundation (NSF) through CAREER grant (Award No. DMR-2337535) and the Materials Research Science and Engineering Center at the University of Michigan (Award No. DMR-2309029). N.M. acknowledges the financial support from the Alexander von Humboldt Foundation. Y.G. and M.Y. acknowledge the support by the Gordon and Betty Moore Foundation's EPiQS Initiative through grant No. GBMF9470. M.X. And W.X. Were supported by the U.S.DOE-BES under Contract DE-SC0023648. L. L. acknowledges the support from the Department of Energy (Award No. DE-SC0020184) for the magnetometry measurements. Y.Z. was supported by the Max Planck partner lab on quantum materials. This work used resources of the Advanced Light Source, which is a DOE Office of Science User Facility under contract no. DE-AC02-05CH11231. The authors acknowledge the University of Michigan College of Engineering for financial support and the Michigan Center for Materials Characterization for use of the instruments and staff assistance. 
 
\section{Author contributions}
S.S. synthesized the single crystals in consultation with N.H.J. S.S., M.X., and A.K.M.A.S. determined the crystallographic orientations. D.Z. and S.S. performed the magnetization measurements in consultation with L.L. and N.H.J. S.S., E.D., E.R., and N.H.J. conducted the ARPES experiment. Y.G., S.S., N.T., and E.R. performed the X-ray microdiffraction measurements and S.S. performed the strain gauge measurements for strain cell calibration in consultation with N.H.J. S.S. prepared the samples for and performed the electrical transport measurements in consultation with N.H.J. and L.L. S.S., N.H.J., L.L., and K.S. analyzed the anomalous Hall data. N.M. performed the DFT calculations in consultation with Y.Z. S.S., N.M., D.Z., M.X., K.S. and N.H.J. wrote the manuscript in consultation with Y.G., A.K.M.A.S., E.D., T.M., M.Y., W.X., C.J., A.B., N.T., E.R., L.L., and Y.Z.

\section{Competing interests}
The Authors declare no Competing Financial or Non-Financial Interests.

\newpage

\clearpage

\clearpage

\clearpage

\clearpage

\title{Strain-tunability of the multipolar Berry curvature in altermagnet MnTe: Supplementary Information}

\author{Shane Smolenski}
\affiliation{Department of Physics, University of Michigan, Ann Arbor, MI 48109, USA}

\author{Ning Mao}
\affiliation{Max Planck Institute for Chemical Physics of Solids, 01187 Dresden, Germany}

\author{Dechen Zhang}
\affiliation{Department of Physics, University of Michigan, Ann Arbor, MI 48109, USA}

\author{Yucheng Guo}
\affiliation{Department of Physics and Astronomy, Rice University, Houston, TX 77005, USA}

\author{A.K.M. Ashiquzzaman Shawon}
\affiliation{Department of Physics, University of Michigan, Ann Arbor, MI 48109, USA}

\author{Mingyu Xu}
\affiliation{Department of Chemistry, Michigan State University, East Lansing, MI 48864, USA}

\author{Eoghan Downey}
\affiliation{Department of Physics, University of Michigan, Ann Arbor, MI 48109, USA}

\author{Trisha Musall}
\affiliation{Department of Physics, University of Michigan, Ann Arbor, MI 48109, USA}

\author{Ming Yi}
\affiliation{Department of Physics and Astronomy, Rice University, Houston, TX 77005, USA}

\author{Weiwei Xie}
\affiliation{Department of Chemistry, Michigan State University, East Lansing, MI 48864, USA}

\author{Chris Jozwiak}
\affiliation{Advanced Light Source, Lawrence Berkeley National Laboratory, Berkeley, CA 94720, USA}

\author{Aaron Bostwick}
\affiliation{Advanced Light Source, Lawrence Berkeley National Laboratory, Berkeley, CA 94720, USA}

\author{Nobumichi Tamura}
\affiliation{Advanced Light Source, Lawrence Berkeley National Laboratory, Berkeley, CA 94720, USA}

\author{Eli Rotenberg}
\affiliation{Advanced Light Source, Lawrence Berkeley National Laboratory, Berkeley, CA 94720, USA}

\author{Lu Li}
\affiliation{Department of Physics, University of Michigan, Ann Arbor, MI 48109, USA}

\author{Kai Sun}
\affiliation{Department of Physics, University of Michigan, Ann Arbor, MI 48109, USA}

\author{Yang Zhang}
\affiliation{Department of Physics and Astronomy, University of Tennessee, Knoxville, TN 37996, USA}
\affiliation{Min H. Kao Department of Electrical Engineering and Computer Science, University of Tennessee, Knoxville, Tennessee 37996, USA}

\author{Na Hyun Jo}
\altaffiliation{nhjo@umich.edu}
\affiliation{Department of Physics, University of Michigan, Ann Arbor, MI 48109, USA}

\renewcommand{\thepage}{S\arabic{page}}  

\def\kill #1{\sout{#1}}
\def\add #1{\textcolor{blue}{#1}} 
\def\addred #1{\textcolor{red}{#1}} 

\newpage

\renewcommand{\thesection}{S\arabic{section}}   
\renewcommand{\thetable}{S\arabic{table}}   
\renewcommand{\figurename}{\textbf{Supplementary Figure}}
\setcounter{figure}{0}
\setcounter{table}{0}

\newpage
\section{Supplementary Note 1: Characterization in the Ambient Condition}
\begin{figure} [ht]
    \includegraphics[width = 6in]{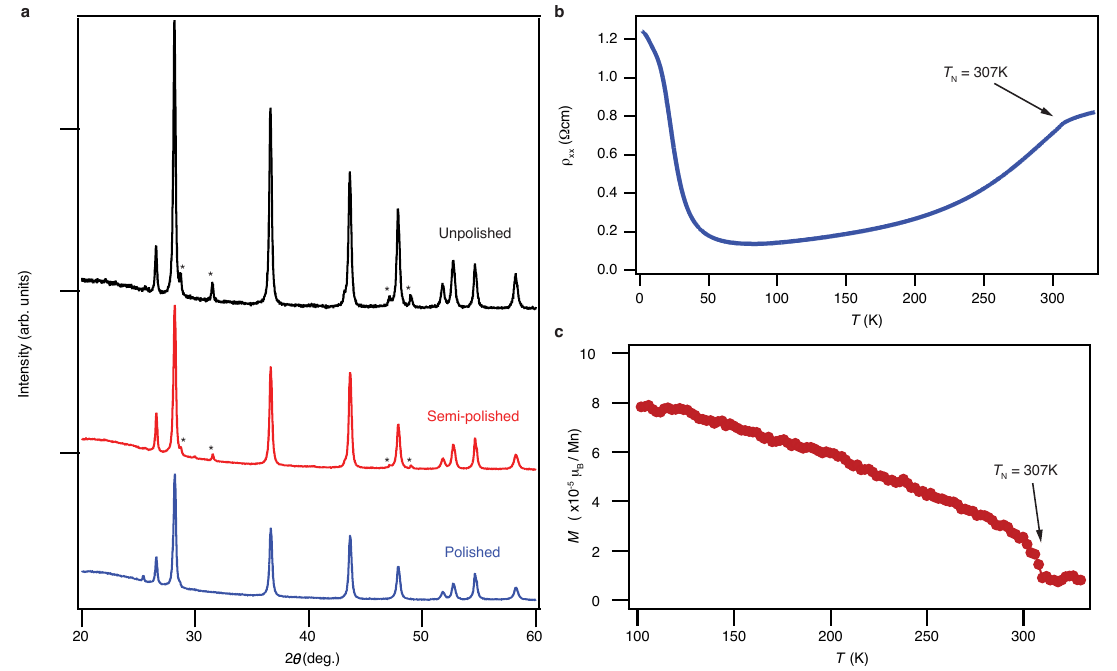}%
    \caption{\textbf{Ambient condition characterization.} 
    \textbf{a}, Powder X-ray diffraction (XRD) spectrum of MnTe single crystals in the unpolished, semi-polished (some, but not all, faces were polished), and polished (all faces were polished) conditions. The asterisks denote peaks originating from MnTe\textsubscript{2} flux on the sample surface that is removed upon surface polishing. \textbf{b}, Resistivity as a function of temperature for an MnTe single crystal for current applied within the \textit{ab}-plane. \textbf{c}, Magnetization along the \textit{c}-axis as a function of temperature measured on a single crystal of MnTe in the temperature range where the AHE was studied. Data was collected while warming in a magnetic field of 10\,Oe along the \textit{c}-axis following field-cooling in a 10\,kOe magnetic field (also applied along the \textit{c}-axis). 
    \label{fig:Ambient}}
\end{figure}

We performed powder X-ray diffraction (XRD) on our single crystals to determine the phase (Supplementary Fig.~\ref{fig:Ambient}a). The recorded spectrum is in excellent agreement with that for hexagonal MnTe with lattice constants $a\,=\,4.149\,\text{\AA}$ and $c\,=\,6.713\,\text{\AA}$ (Supplementary Fig.~\ref{fig:Ambient}a). In addition to the majority MnTe XRD peaks, we observe some small contributions from MnTe\textsubscript{2}. These peaks disappeared after polishing the surfaces of the crystals prior to powderization (Supplementary Fig.~\ref{fig:Ambient}a), suggesting that MnTe\textsubscript{2} only arises from flux coating the crystal surface while the bulk of the crystals are pure MnTe. All samples were fully polished prior to electrical transport measurements.

We next performed temperature-dependent resistivity measurements on a crystal mounted on a regular Quantum Design Physical Properties Measurement System (PPMS) Electrical Transport Option (ETO) puck (direction of the electrical current was randomly oriented within the \textit{ab}-plane). The resistivity behavior as a function of temperature, marked by decreasing resistivity with deceasing temperature above 50\,K with a steep increase in resistivity as the temperature is lowered below 50\,K, is in good agreement with previous reports~\cite{KLUCZYK2024, UCHIDA1956, SQUIRE1939,AOYAMA2024} (Supplementary Fig.~\ref{fig:Ambient}b). A kink in the resistivity curve, which has been reported to occur at the Néel temperature $T_{\textrm{N}}$~\cite{KLUCZYK2024}, appears at 307\,K, also in good agreement with previous reports~\cite{KLUCZYK2024, SQUIRE1939, UCHIDA1956,WU2025,BARAL2023}. 

To characterize the magnetic properties of our samples, we performed temperature-dependent magnetization measurements on a bulk sample using the vibrating sample magnetometry (VSM) option in the PPMS (Supplementary Fig.~\ref{fig:Ambient}c). The sample was cooled in a magnetic field of 10\,kOe to 75\,K and the measurement was performed upon warming with a set field of 10\,Oe. Both the cooling and measurement fields were applied along the \textit{c}-axis. Below a transition temperature of $T_{\textrm{N}}=307\,K$, the magnetization steadily increases, reaching a value of roughly $8\times10^{-5} \,\mu_B$ per Mn atom at 100\,K. The transition is in excellent agreement with the kink observed in our temperature-dependent resistivity measurements and published results for $T_{\textrm{N}}$~\cite{KLUCZYK2024, SQUIRE1939, UCHIDA1956,WU2025,BARAL2023}. The non-zero magnetization below $T_{\textrm{N}}$ is the expected signature of weak ferromagnetism (WF) that is allowed by the symmetry of MnTe and is in agreement, in both shape and magnitude, with previous reports~\cite{KLUCZYK2024}. While there is a slight net magnetization in MnTe, we emphasize that the magnitude is exceedingly small in comparison with ferromagnetic materials and, as argued in the main text, not the driving source of the anomalous Hall effect (AHE) in MnTe.

We explored the electronic structure of MnTe using angle-resolved photoemission spectroscopy (ARPES). All ARPES measurements were performed at Beamline 7.0.2 (MAESTRO) at the Advanced Light Source at Lawrence Berkeley National Laboratory. Linear horizontal (\textit{p}-polarized) light was used and the sample temperature was set to 25\,K. The Brillouin zone of MnTe is shown for reference in Supplementary Fig.~\ref{fig:ARPES}a. We first identified the photon energies associated with the high-symmetry points $\Gamma$ and A through a photon energy-dependent measurement. By focusing on a highly dispersive band that reaches its minimum at the $A$ point 2.5\,eV below the Fermi level $E_{\text{F}}$~\cite{OSUMI2024}, we identified the photon energies associated with the $A$ and $\Gamma$ points (Supplementary Fig.~\ref{fig:ARPES}b). We observe the characteristic 6-fold symmetry within the $\Gamma$MK plane (Supplementary Fig.~\ref{fig:ARPES}c), where the $\Gamma$, M, and K points can be clearly determined. The 6-fold pattern with intensity maxima between $\Gamma$ and K is in good agreement with previous ARPES measurements~\cite{KREMPASKY2024,LEE2024,OSUMI2024}. We then measured the band dispersion along the high symmetry $\Gamma$-K line (Supplementary Figs.~\ref{fig:ARPES}d,e). The highest-energy valence band can be identified and its shape is in good qualitative agreement with our DFT calculations (Main Text Fig. 1b). By taking the second derivative of the momentum distribution curves, we identify the band maximum (intensity edge) along the $\Gamma$-K  path to be $\sim 70 \pm 20$\,meV below $E_{\text{F}}$ (Supplementary Fig.~\ref{fig:ARPES}e), where the uncertainty arises from experimental band broadening and slightly imperfect alignment within the Brillouin zone. The position of $E_{\text{F}}$ indicates a natural \textit{p} doping of our samples, as is common in the literature~\cite{LEE2024,KREMPASKY2024} and in agreement with our Hall measurements (Supplementary Fig.~\ref{fig:AHE_Extraction}a).

\begin{figure} [ht]
    \includegraphics[width = 6in]{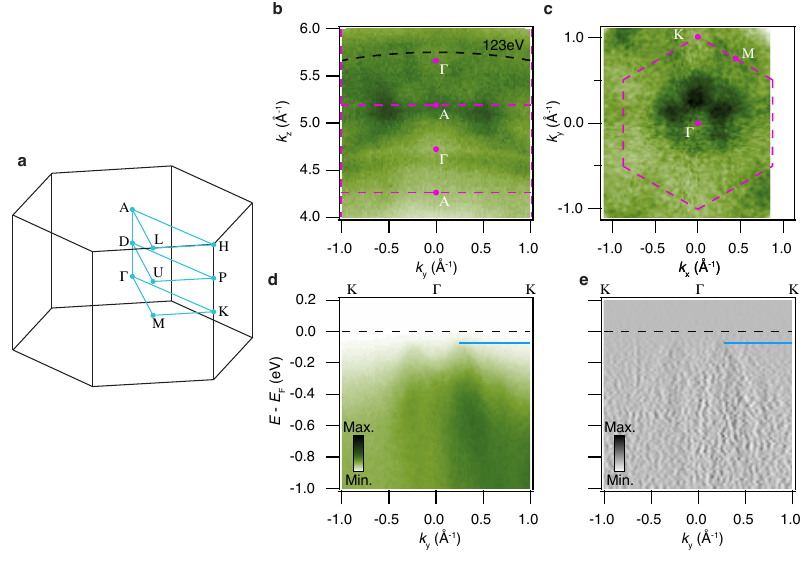}%
    \caption{\textbf{ARPES measurements of MnTe.} 
    \textbf{a}, Brillouin zone of MnTe with high symmetry points labeled. \textbf{b}, Iso-energy plot in the $k_y$-$k_z$ plane taken at a binding energy of 2.5\,eV. An inner potential of $V_0\,=\,7.2$\,eV was used to convert photon energy to $k_z$ within the final state nearly-free electron approximation. \textbf{c}, Iso-energy plot in the $k_x$-$k_y$ plane taken at a binding energy of 70\,meV. A photon energy of $h\nu=120$\,eV was used. \textbf{d}, Energy dispersion along the $\Gamma$-K high symmetry path using a photon energy of 123\,eV. \textbf{e}, Second derivative of (\textbf{d}), demonstrating the band maximum to be $\sim 70 \pm 20$\,meV below $E_{\text{F}}$. The blue line in (\textbf{d}),(\textbf{e}) highlights the band maximum along the $\Gamma$-K high symmetry path.
    \label{fig:ARPES}}
\end{figure}

\clearpage
\section{Supplementary Note 2: Extraction of Anomalous Hall Resistivity}
\begin{figure} [ht]
    \includegraphics[width = 6in]{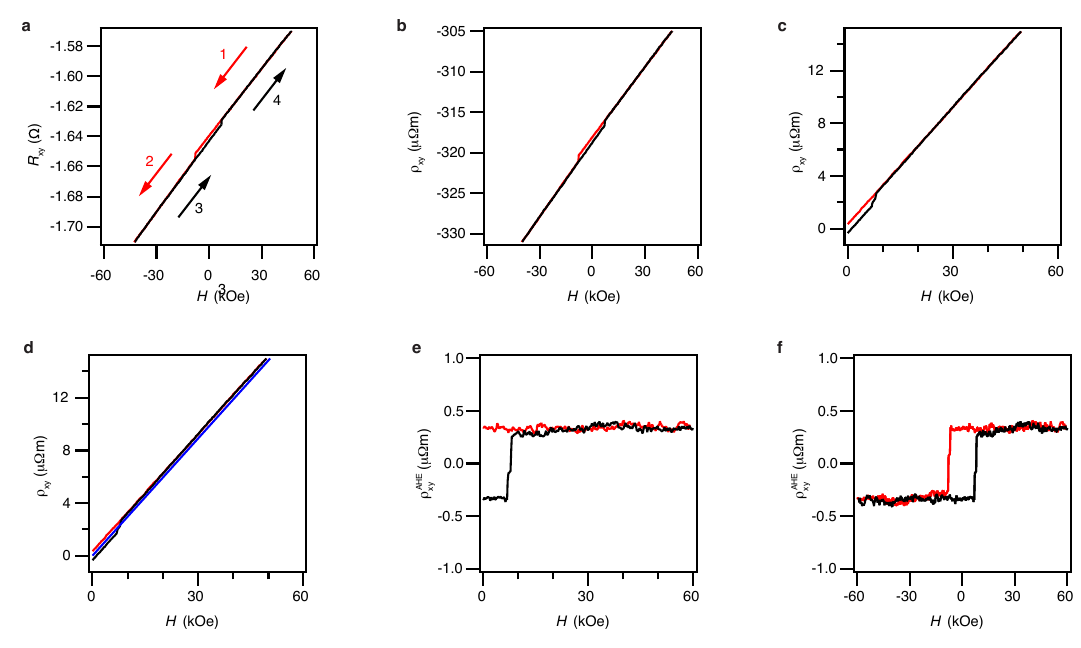}%
    \caption{\textbf{Extraction of AHE from transverse resistance measurements.} 
    \textbf{a}, Transverse resistance $R_{xy}$ as measured during the experiment with quadrants labeled. Arrows show the magnetic field sweep direction. All data is shown for Sample A in the Mounted condition at 200\,K. \textbf{b}, Transverse resistivity, achieved by multiplying (\textbf{a}) by the measured thickness. \textbf{c}, Anti-symmetrized Hall resistivity $\rho_{xy}$, achieved using Equations~\ref{Rho_inc} (Black curve) and \ref{Rho_dec} (red curve). \textbf{d}, Same as (\textbf{c}), but with the linear fit to $\rho_{xy}^{\text{Dec}}$ overlaid (blue curve) and offset to pass through the origin. \textbf{e}, Anomalous Hall resistivity $\rho_{xy}^\text{AHE}$ from $H=0$\,kOe to $H=60$\,kOe, achieved by subtracting the linear fit shown in (\textbf{d}). \textbf{f}, Same as (\textbf{e}), but from $H=-60$\,kOe to $H=60$\,kOe. The region $H=-60$\,kOe to $H=0$\,kOe is found from using the fact that $\rho_{xy}^{\text{AHE}}(+H)= -\rho_{xy}^{\text{AHE}}(-H)$. 
   \label{fig:AHE_Extraction}}
\end{figure}

The AHE measurements were performed by measuring the transverse voltage under a longitudinal electric field and vertical magnetic field from +60\,kOe to 0\,kOe (Quadrant 1), 0\,kOe to -60\,kOe (Quadrant 2), -60\,kOe to 0\,kOe (Quadrant 3), and 0\,kOe to +60\,kOe (Quadrant 4) (Supplementary Fig.~\ref{fig:AHE_Extraction}a). Since the measured resistance includes contributions from the ordinary Hall effect, longitudinal magnetoresistance (MR), and the AHE, it is essential to carefully extract the AHE component from the measured signal. First, transverse resistance $R_{xy}$ was converted to transverse resistivity $\rho_{xy}$ by multiplying by the sample thickness $\sim200\,\mu m$ (Supplementary Fig.~\ref{fig:AHE_Extraction}b), which was measured with an optical microscope. Then, to remove unwanted contributions from longitudinal MR, the data was antisymmetrized. In order to not wash out the AHE contribution, the antisymmetrization was done according to:
\begin{equation}
    \rho_{xy}^{\text{Inc}}(H) = \frac{\rho_{xy,4}(+H)-\rho_{xy,2}(-H)}{2}
    \label{Rho_inc}
\end{equation}
\begin{equation}
    \rho_{xy}^{\text{Dec}}(H) = \frac{\rho_{xy, 1}(+H)-\rho_{xy,3}(-H)}{2},
    \label{Rho_dec}
\end{equation}
where  $\rho_{xy}^{\text{Inc}}$ is the Hall resistivity for fields of increasing magnitude, $\rho_{xy}^{\text{Dec}}$ is the Hall resistivity for fields of decreasing magnitude, and $\rho_{xy,i}$ is the transverse resistivity recorded in Quadrant \textit{i} (Supplementary Fig.~\ref{fig:AHE_Extraction}c)~\cite{BALICAS2011}. Both $\rho_{xy}^{\text{Inc}}$ and $\rho_{xy}^{\text{Dec}}$ contain contributions from the ordinary Hall effect and AHE. To remove the ordinary Hall component, $\rho_{xy}^{\text{Dec}}$ was linearly fit and the resulting fit was then offset so as to pass through the origin (Supplementary Fig.~\ref{fig:AHE_Extraction}d). This offset fit was then subtracted from both $\rho_{xy}^{\text{Inc}}$ and $\rho_{xy}^{\text{Dec}}$, resulting in $\rho_{xy}^{\text{AHE, Inc}}$ and $\rho_{xy}^{\text{AHE, Dec}}$, both of which make up the total anomalous Hall resistivity $\rho_{xy}^{\text{AHE}}$ for magnetic fields from 0 to 60\,kOe (Supplementary Fig.~\ref{fig:AHE_Extraction}e). Finally, since the AHE is odd under time reversal, $\rho_{xy}^{\text{AHE}}(+H)= -\rho_{xy}^{\text{AHE}}(-H)$, $\rho_{xy}^{\text{AHE}}$ was copied, multiplied by -1, and reflected across $H=0$ to produce $\rho_{xy}^{AHE}$ from -60\,kOe to +60\,kOe (Supplementary Fig.~\ref{fig:AHE_Extraction}f) .

\clearpage
\section{Supplementary Note 3: Strain Determination}
The total strain applied to the sample $\varepsilon^{tot}$ is given by the sum of thermal strain $\varepsilon^T$ and mechanical strain $\varepsilon^M$, $\varepsilon^{tot}=\varepsilon^T+\varepsilon^M$. Therefore, it is essential to quantify both thermal and mechanical strain to determine the total strain experienced by the sample.

\subsection{Thermal Strain}
Thermal strain is generated by the relative difference in thermal contraction between that of MnTe and the gap on the stress cell over which the sample is mounted (hereon referred to as the gap). Because the sample is fixed in place over the gap, the length of the gap determines the sample length. Thus, if MnTe naturally contracts more (less) than the gap, an effective tensile (compressive) stress is generated on the sample that varies as a function of temperature. This can be described as: 
\begin{equation}
    \begin{aligned}
    \varepsilon^{T}(T) & = \frac{L_{\text{Gap}}(T)-L_{\text{MnTe}}(T)}{L_{\text{MnTe}}(T)}\\&=\frac{L(293\,\text{K})\times[(1+\varepsilon_{\text{Gap}}(T))-(1+\varepsilon_{\text{MnTe}}(T))]}{L(293\,\text{K})\times(1+\varepsilon_{\text{MnTe}}(T))}\\&=\frac{\varepsilon_{\text{Gap}}(T) - \varepsilon_{\text{MnTe}}(T)}{1+\varepsilon_{\text{MnTe}}(T)},
    \end{aligned}
    \label{thermal_strain}
\end{equation}
where $L_{\text{Gap}}(T)$ is the temperature-dependent length of the gap, $L_{\text{MnTe}}(T)$ is the natural temperature-dependent length of the MnTe sample if it were free to contract, $L(293\,\text{K})$ is the length of the gap/sample at 293\,K, $\varepsilon_{\text{Gap}}(T)=\frac{L_{\text{Gap}}(T)-L(293\,\text{K})}{L(293\,\text{K})}$, and $\varepsilon_{\text{MnTe}}(T)=\frac{L_{\text{MnTe}}(T)-L(293\,\text{K})}{L(293\,\text{K})}$. Thus, both the thermal contraction of the gap and MnTe must be known in order to calculate the effective thermal strain.

The thermal contraction of the gap was measured using a Micro-Measurements C5K-06-S5198-350-33F strain gauge. The gauge was mounted over the gap using Loctite EA1C in the same manner as our MnTe samples were mounted for transport measurements. The gauge was then cooled to 2\,K and the gauge resistance was measured as a function of temperature from 2\,K to 330\,K at a rate of 1\,K per minute in the PPMS (red curve in Supplementary Fig.~\ref{fig:Thermal_Strain_Determination}b). The strain was calculated from the gauge resistance using $\varepsilon=\frac{1}{GF}\times\frac{R(T)-R(293\,\text{K})}{R(293\,\text{K})},$ where $GF$ is the gauge factor provided by the manufacturer and $R(T)$ is the measured temperature-dependent resistance of the gauge mounted across the gap. However, this measurement contains both the change in length of the gap (“geometric effects”) and the intrinsic changes in the gauge resistance as a function of temperature unrelated to the physical change in length of the gauge (“thermal effects”). Only the geometric effects describe the change in length of the gap as a function of temperature and thus the thermal effects must be measured and subtracted. To do so, a separate gauge was mounted on high purity 101 Cu using Loctite EA1C. The gauge resistance was then measured from 2\,K to 330\,K in the PPMS at a rate of 1\,K per minute. The resultant curve was then converted to relative contraction (relative to the length at 293\,K) and compared with the known contraction of Cu reported in the literature~\cite{CORRUCCINI1961}. The difference between these curves is taken as the thermal effects of the strain gauge (Supplementary Fig.~\ref{fig:Thermal_Strain_Determination}a). This was then subtracted from the gauge measurement over the gap to produce the thermal contraction of the gap as a function of temperature (Supplementary Fig.~\ref{fig:Thermal_Strain_Determination}b).

The thermal contraction of the MnTe lattice constants as a function of temperature has been previously reported in the literature and was used to generate the relative contraction (relative to 293\,K) of MnTe~\cite{BARAL2023}. Thus, with the relative thermal contraction of both MnTe and the gap measured, the effective thermal strain as a function of temperature could then be calculated using Eq.~\ref{thermal_strain}. The relative thermal contraction of MnTe, the gap, and the effective thermal strain are all plotted in Main Text Fig. 2b.

\begin{figure} [ht]
    \includegraphics[width = 6in]{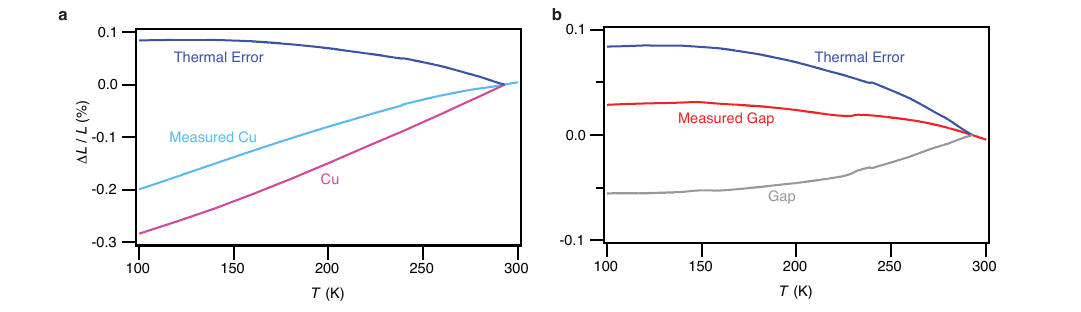}%
    \caption{\textbf{Determination of thermal strain.} 
    \textbf{a}, Relative change in length of Cu as measured by the strain gauge (light blue curve) and reported literature values (pink curve). The difference between the two results in the thermal effects of the gauge (dark blue curve). \textbf{b}, Relative change in length of the gap measured using the strain gauge (red curve),  the determined thermal effects from (\textbf{a}) (dark blue curve), and the true temperature-dependent relative change in gap length (gray curve) obtained by subtracting the thermal effects from the measured gap.     \label{fig:Thermal_Strain_Determination}}
\end{figure}

\subsection{Mechanical Strain}

\begin{figure} [ht]
    \includegraphics[width = 3in]{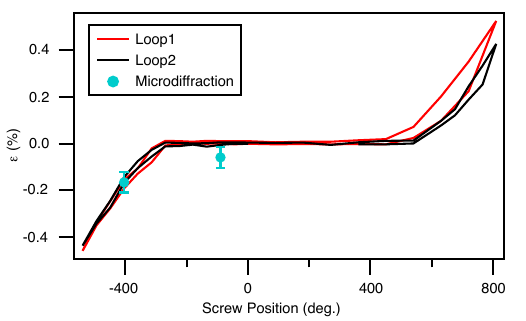}%
    \caption{\textbf{Determination of mechanical strain.} 
    Relative change in length of the gap as a function of screw position as measured by the strain gauge during two separate angular sweeps (red and black curves) and X-ray microdiffraction (light blue markers).
    \label{fig:Strain_Gauge_Calibration}}
\end{figure}

The other contribution to total strain is mechanical strain $\varepsilon^M$, which is the strain generated by mechanically moving the plates of the stress cell via turning of the screw on the cell. To measure the mechanical strain, the resistance of a strain gauge (the same gauge mounted over the gap of the stress cell used for the thermal strain measurements) was measured as a function of screw position at room temperature ($T\approx 293\,K$). The gauge resistance was measured by turning the screw from relative position of $0\degree$ to $-540\degree$ to $+810\degree$ to $0\degree$ (negative angle refers to clockwise rotation) and the relative change in length was calculated as $\varepsilon=\frac{1}{GF}\times\frac{R(\Theta)-R(0\degree)}{R(0\degree)},$ where $GF$ is the gauge factor given by the manufacturer, and $R(\Theta)$ is the measured gauge resistance as a function of screw angle $\Theta$. As expected, clockwise rotation (negative angles) corresponds to compressive strain and counterclockwise rotation (positive angles) corresponds to tensile strain. The region between $\sim -270\degree$ and $\sim 500\degree$ where no strain was generated is a result of the stress cell design and is thus expected. Two cycles were performed in good agreement with each other (Supplementary Fig.~\ref{fig:Strain_Gauge_Calibration}).

\begin{figure} [ht]
    \includegraphics[width = 6in]{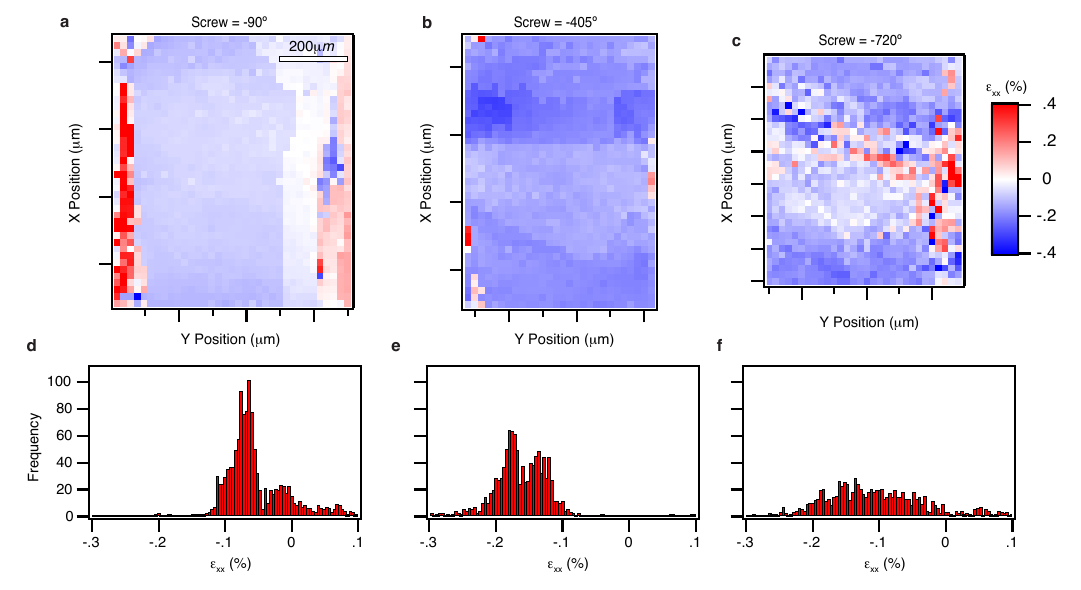}%
    \caption{\textbf{Microdiffraction of strained MnTe.} \textbf{a}-\textbf{c}, Spatial maps of the longitudinal strain (i.e. along the direction of applied stress) experienced by a sample of MnTe mounted across the gap of the stress cell at screw positions of -90\degree (\textbf{a}), -405\degree (\textbf{b}), and -720\degree (\textbf{c}). \textbf{d}-\textbf{f}, Histograms of the longitudinal strain values for screw position -90\degree (\textbf{d}), -405\degree (\textbf{e}), and -720\degree (\textbf{f}). The applied strain is relatively homogeneous until fracture at -720\degree and agrees well with changes to the gap measured with the strain gauge. Stress was applied along the $x$-direction in this coordinate system.
    \label{fig:Microdiffraction}}
\end{figure}

While each MnTe sample is mounted firmly in epoxy over the gap, it is possible that not all of the strain (relative change in length) experienced by the gap is transferred to the sample. To explore how the strain is experienced by an MnTe sample mounted over the gap itself, we performed X-ray Laue microdiffraction measurements at Beamline 12.3.2 at the Advanced Light Source at Lawrence Berkeley National Laboratory. The X-ray beam (spot size of $1\,\mu \text{m}^2$) was rastered over a shaped sample of MnTe (typical dimensions of samples used for transport measurements) mounted over the gap of the stress cell and the X-ray diffraction pattern was measured by a 2D detector at each step. The diffraction patterns enable refinement of the local lattice parameters, from which the spatially resolved deviatoric strain can be calculated with respect to the known MnTe crystal structure. Measurements were performed at three separate screw positions; the spatial maps of the deviatoric strain along the motion of the plates ($x$-direction in our Cartesian coordinate system) as well as the histograms of the strain distribution are shown in Supplementary Fig.~\ref{fig:Microdiffraction}. Averaged compressive strains of $-0.059\pm0.046\%$ and $-0.166\pm0.44\%$ were experienced by the sample at screw positions of $-90\degree$ and $-405\degree$, respectively, where the error bars represent the standard deviations. At a screw position of $-720\degree$, the sample fractured (Supplementary Fig.~\ref{fig:Microdiffraction}c). These data points are plotted in Supplementary Fig.~\ref{fig:Strain_Gauge_Calibration} with closed blue circles, demonstrating agreement with the strain gauge measurements. Therefore, we find that the relative change in gap size is effectively transferred to the sample in a homogeneous manner and that the mechanical strain measured by the strain gauge is an accurate estimate within $\sim0.045\%$ of the strain experienced by the sample.

\subsection{Total Strain}
The thermal and mechanical strains measured in the above sections are only for strain along the direction of motion of the plates (defined as the $x$-direction in our Cartesian coordinate system). However, since a uniaxial stress is being physically applied, we may expect other strain components to be generated as well. The applied stress and other strain components may be calculated using the elastic tensor $C$ as $\mathbf{S}=C\mathbf{\varepsilon},$ where $\mathbf{S}$ is the stress tensor and $\mathbf{\varepsilon}$ is the strain tensor. We used the calculated elastic tensor of MnTe from the Materials Project~\cite{JAIN2013,DEJONG2015}. Thus, expanded, we may write:
\begin{equation}
    \begin{bmatrix}
        S_{xx}\\ S_{yy} \\ S_{zz} \\ S_{yz} \\ S_{xz} \\S_{xy}
    \end{bmatrix}
    =
    \begin{bmatrix}
        80 & 38 & 32 & 0 & 0 & 0 \\
        38 & 80 & 32 & 0 & 0 & 0 \\
        32 & 32 & 61 & 0 & 0 & 0 \\
        0 & 0 & 0 & 28 & 0 & 0 \\
        0 & 0 & 0 & 0 & 28 & 0 \\
        0 & 0 & 0 & 0 & 0 & 21 \\
    \end{bmatrix}
    \begin{bmatrix}
        \varepsilon_{xx}\\ \varepsilon_{yy} \\ \varepsilon_{zz} \\ 2\varepsilon_{yz} \\ 2\varepsilon_{xz} \\2\varepsilon_{xy}
    \end{bmatrix},
\end{equation}
where the components of $C$ are in units of GPa and . For an MnTe sample mounted with the length along $[10\bar10]$ ($x$-direction in our defined Cartesian coordinate system), stress is only applied along the $x$-direction, leaving $S_{xx}$ as the only non-zero component of the stress tensor. Thus, $\varepsilon_{xx}$, $\varepsilon_{yy}$, and $\varepsilon_{zz}$ are the only non-zero components of the strain tensor. This simplifies the equation to:

\begin{equation}
    \begin{bmatrix}
        S_{xx}\\ 0 \\ 0 \\ 0 \\ 0 \\0
    \end{bmatrix}
    =
    \begin{bmatrix}
        80 & 38 & 32 & 0 & 0 & 0 \\
        38 & 80 & 32 & 0 & 0 & 0 \\
        32 & 32 & 61 & 0 & 0 & 0 \\
        0 & 0 & 0 & 28 & 0 & 0 \\
        0 & 0 & 0 & 0 & 28 & 0 \\
        0 & 0 & 0 & 0 & 0 & 21 \\
    \end{bmatrix}
    \begin{bmatrix}
        \varepsilon_{xx}\\ \varepsilon_{yy} \\ \varepsilon_{zz} \\ 0 \\ 0 \\0
    \end{bmatrix}.
\end{equation}

With $\varepsilon_{xx}$ known from our above thermal and mechanical calibrations, $S_{xx}$, $\varepsilon_{yy}$, and $\varepsilon_{zz}$ can then be calculated.

\newpage
\section{Supplementary Note 4: Reversibility of the Applied Strain-Induced Changes to the AHE}
We studied the reversibility of the strain-induced changes to further explore the connection between the sign of the AHE and strain. To do so, after performing the Hall measurements in the Compressed condition (-0.118\% mechanical strain) on Sample A, we released the mechanical strain (turned the screw back to its original position of $0\degree$) and repeated the Hall measurements at various temperatures. The saturated anomalous Hall resistivity $\rho_{\text{sat}}$ as a function of temperature is shown in Supplementary Fig.~\ref{fig:Reversibility}a for Sample A in the Mounted condition before compression, in the Compressed condition, and again in the Mounted condition after releasing the mechanical compressive stress. The AHE behavior in the Mounted condition after compression largely reverts back to that of the original Mounted condition before compression. For example, at 220\,K, the sign of $\rho_{\text{sat}}$ is positive both before the applied compressive stress and after its release while the sign is negative under compression. Moreover, the transition temperature $T^*$ originally occurs at $\sim225$\,K in the Mounted condition, drops to $\sim190$\,K in the Compressed condition, and then returns to $\sim235$\,K upon release of the compressive stress (Supplementary Fig~\ref{fig:Reversibility}b). The slight difference between $T^*$ for the Mounted condition before and after compression may be explained by imperfect strain release by the cell or from changes in the sample developed as a result of strain (e.g. domain alignment). We thus demonstrate that the strain-induced changes to the AHE are reversible and further support the connection between the AHE reversal and strain. 

\begin{figure} [ht]
    \includegraphics[width = 6in]{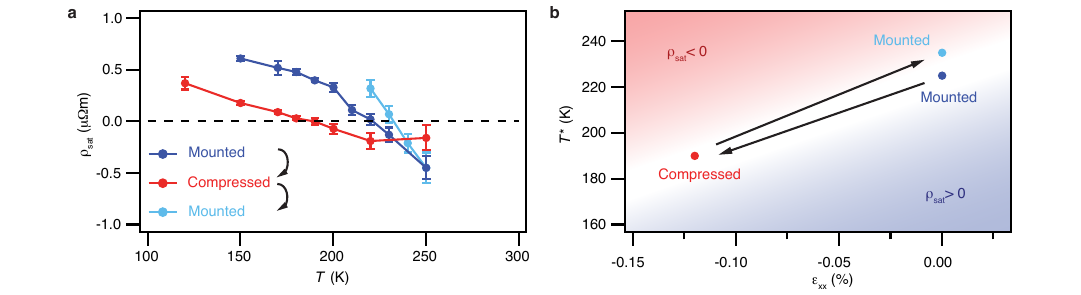}%
    \caption{\textbf{Reversibility of the strain-induced changes to the AHE in Sample A.} 
    \textbf{a}, $\rho \textsubscript{sat}$ as a function of temperature for the Mounted condition before compression (dark blue trace), Compressed condition (red trace), and Mounted condition after compression (light blue trace). \textbf{b}, Transition temperature $T^*$ as a function of applied strain.
    \label{fig:Reversibility}}
\end{figure}

\clearpage
\section{Supplementary Note 5: Magnetization Measurements Under Strain}

\begin{figure} [ht]
    \includegraphics[width = 6in]{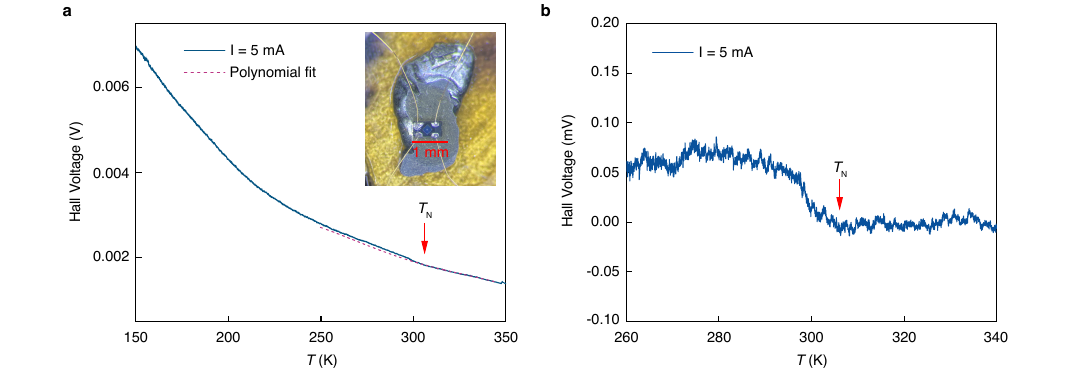}%
    \caption{\textbf{Schematic of the Hall sensor setup and temperature-dependent magnetization measurement.} 
    \textbf{a}, Hall voltage of the Hall sensor as a function of temperature. A direct current of 5\,mA was applied to the Hall sensor to measure the Hall voltage. The red short-dashed line represents a polynomial fit to the high-temperature paramagnetic background signal detected by the Hall sensor. Upon warming from 100\,K to 350\,K, a clear anomaly near 306\,K corresponding to the Néel transition was observed. The insert shows the schematic of the Hall sensor setup. \textbf{b}, Data after subtraction of the polynomial background. 
    }
    \label{fig:strain_mag}
\end{figure}

To characterize possible changes in magnetization under applied strain, we used a Hall sensor (AKM HQ-0A11) to measure the temperature dependence of the magnetization of MnTe samples both with and without strain. A direct current of 5\,mA was applied to the Hall sensor, and the Hall voltage was recorded. For the free-standing bulk sample ($\sim$1.4\,mm × $\sim$1.5\,mm x $\sim$0.5\,mm), the sample was first cooled in a field of approximately 10\,kOe to a low temperature (below 100\,K). The large external field was then removed and replaced by a small field of 30\,Oe. The sample was warmed from 100\,K to 350\,K to obtain the temperature-dependent magnetization. In this configuration, we observed a transition near 306\,K (Supplementary Fig.~\ref{fig:strain_mag}), consistent with the Néel temperature determined from VSM measurements (Supplementary Fig.~\ref{fig:Ambient}c) and previous reports.

For measurements under strain, an MnTe crystal was shaped and mounted on the strain cell, resulting in a smaller sample size ($\sim$1.2\,mm × $\sim$0.5\,mm × $\sim$0.2\,mm). The same measurement protocol was followed for both the unstrained and strained states. In both cases, no sharp magnetization transition could be resolved within our sensitivity limit (corresponding to a magnetic field resolution better than 100\,nT at $I=$\,5mA), although a small net magnetization was present. This absence of a clear transition signal in the strained sample is attributed to the extremely small ferromagnetic component in MnTe, combined with the reduced sample volume and the limited sensitivity of the Hall sensor setup compared to VSM.

\clearpage
\section{Supplementary Note 6: Strain-induced Reversal of the Anomalous Hall Resistivity at Zero Magnetic Field}
From our observation that the sign of the AHE hysteresis reverses under strain, we know that the product of $M_z$ and $\sigma_{xy}$ changes sign with strain. However, it is not immediately clear from our saturated field measurements whether it is $M_z$ or $\sigma_{xy}$ that changes sign. Importantly, these two pictures can be clearly separated by measuring the anomalous Hall resisitvity $\rho_{xy}^{\text{AHE}}$ at zero field. For a system in a polarized state (i.e. all domains have same $\bf{L}$ / $M_z$), a reversal of $M_z$ (that doesn’t reverse $\bf{L}$), would not change the sign of $\sigma_{xy}$ (and $\rho_{xy}^{\text{AHE}}$, relatedly), assuming small values of $M_z$ and that $\sigma_{xy}^{\text{WF}} \approx 0$ holds, as argued in the Main Text. However, a direct tuning of the sign of $\sigma_{xy}^{\text{AM}} \approx \sigma_{xy}$ would change the sign of the anomalous Hall conductivity at zero field.

To explore this, we measured $\rho_{xy}^{\text{AHE}}$ for $H_z=0$ on a sample in the Mounted condition at two separate temperatures (strain values), one above the AHE hysteresis reversal critical temperature $T^*$, $T > T^*$ (below AHE hysteresis reversal critical strain $\varepsilon_{xx}^*$, $\varepsilon_{xx} < \varepsilon_{xx}^*$), and one for $T < T^*$ ($\varepsilon_{xx} > \varepsilon_{xx}^*$). First, at $T > T^*$ ($\varepsilon_{xx} < \varepsilon_{xx}^*$), we polarized the system into the +$M_z$ state by ramping $H_z$ from -60\,kOe to 60kOe (black curve in Supplementary Fig.~\ref{fig:Zero_Field}a). This stabilized $\rho_{xy}^{\text{AHE}} < 0$ for $H_z > 0$. The field was then reduced from 60\,kOe to 0\,kOe where the +$M_z$ state with $\rho_{xy}^{\text{AHE}} < 0$ was preserved (red curve in Supplementary Fig.~\ref{fig:Zero_Field}a). Then, with $H_z=0$, the temperature was lowered so that $T < T^*$ ($\varepsilon_{xx} > \varepsilon_{xx}^*$) and the AHE was measured from 0\,kOe to 60\,kOe (blue curve in Supplementary Fig.~\ref{fig:Zero_Field}b). As shown in Supplementary Fig.~\ref{fig:Zero_Field}b, $\rho_{xy}^{\text{AHE}} > 0$ for $H_z > 0$ starting immediately from $H_z=0$. This demonstrates that the sign of $\sigma_{xy}^{\text{AM}}$, and not $M_z$, flips with strain, driving the hysteresis reversal of the AHE. Therefore, we can conclude that strain-induced changes to the Berry curvature, rather than magnetization, are responsible for the sign reversal of the AHE hysteresis.

 \begin{figure} [ht]
    \includegraphics[width = 3in]{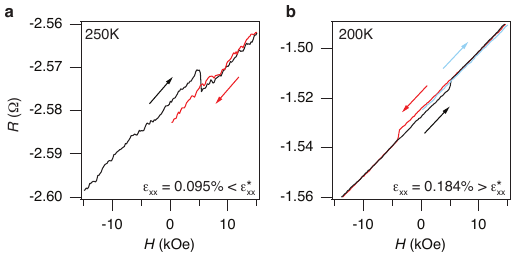}%
    \caption{\textbf{Measurement of the Anomalous Hall Effect at Zero Magnetic Field.} 
    \textbf{a}, Hall resistance as a function of magnetic field at 250\,K ($T>T^*$, $\varepsilon_{xx}<\varepsilon_{xx}^*$). The magnetic field was swept from -60\,kOe to +60\,kOe (black curve) and then from +60\,kOe to 0\,kOe (red curve) to polarize the magnetic state. \textbf{b}, Hall resistance as a function of magnetic field at 200\,K ($T<T^*$, $\varepsilon_{xx}>\varepsilon_{xx}^*$) after cooling below $T^*$ in zero magnetic field. The Hall resistance was measured as the field was swept from 0\,kOe to +60\,kOe (light blue curve), +60\,kOe to -60\,kOe (red curve), and from -60\,kOe to +60\,kOe (black curve). The arrows indicate the direction of field ramping for each measurement.
    \label{fig:Zero_Field}
    }
\end{figure}

\clearpage
\section{Supplementary Note 7: Orientation Determination}
Due to the morphology of our MnTe crystals, only $(0001)$ can be identified visually (as the flat faces of the plate-like crystals). To determine the crystallographic axes within the \textit{ab}-plane, we used single crystal X-ray diffraction (XRD) and diffraction contrast tomography (DCT). 

\begin{figure} [ht]
    \includegraphics[width = 6in]{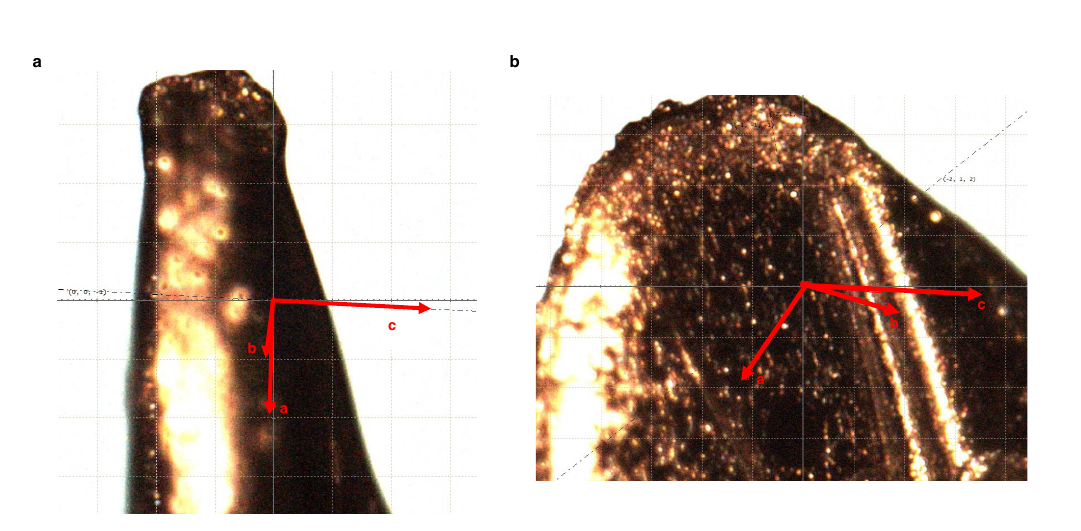}%
    \caption{\textbf{Orientation determination via single crystal XRD.} 
    \textbf{a},\textbf{b}, Optical microscope images of an MnTe single crystal mounted for X-ray diffraction measurements. Red arrows indicate the crystallographic axes determined by face indexing.
   \label{fig:SCXRD}}
\end{figure}

For single crystal XRD, an MnTe single crystal was selected for analysis (Supplementary Fig.~\ref{fig:SCXRD}). This crystal was affixed to a glass fiber with superglue, facilitating its examination via a Rigaku XtalLAB Synergy, Dualflex, Hypix single crystal X-ray diffractometer. The apparatus was operated at room temperature. Crystallographic data acquisition was conducted employing $\omega$ scan methodology, utilizing Mo K$\alpha$ radiation ($\lambda$ = 0.71073\,\AA{}) emitted from a micro-focus sealed X-ray tube under operating conditions of 50\,kV and 1\,mA. The determination of the experimental parameters, including the total number of runs and images, was derived algorithmically from the strategy computations facilitated by the CrysAlisPro software, version 1.171.42.101a (Rigaku OD, 2023). Subsequent data reduction processes incorporated corrections for Lorentz and polarization effects. 

\begin{figure} [ht]
    \includegraphics[width = 6in]{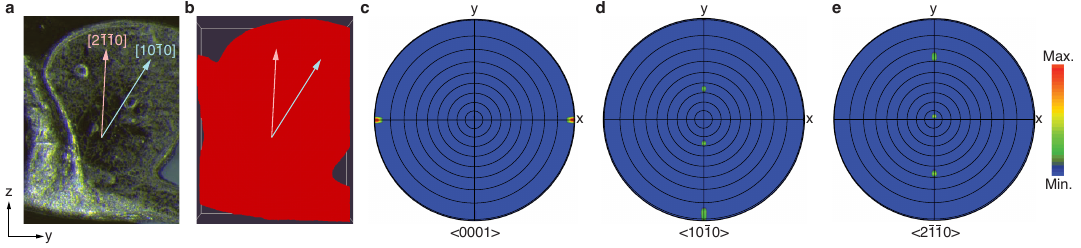}%
    \caption{\textbf{Orientation determination via diffraction contrast tomography.} 
    \textbf{a}, Optical image of an MnTe sample. \textbf{b}, Virtually reconstructed image of the MnTe sample from the DCT measurement. \textbf{c}-\textbf{e}, Pole figures indicating the directions of $<\!0001\!>$ (\textbf{c}), $<\!10\bar10\!>$ (\textbf{d}), and $<\!2\bar1\bar10\!>$ (\textbf{e}). Each concentric ring from the origin of the pole figures represents the angle from the \textit{z}-axis in steps of $10\degree$.
   \label{fig:DCT}}
\end{figure}

For DCT, large crystals with a clearly recognizable $(0001)$ face were selected and mounted on a rotating post such that the $[0001]$ was directed towards the X-ray beam (Supplementary Fig.~\ref{fig:DCT}a). The sample was then rotated $360\degree$ about an axis perpendicular to the beam where the absorption was measured at each step; this enabled the creation of a 3D model of the crystal(Supplementary Fig.~\ref{fig:DCT}b). The measurement was then repeated, but with the Laue diffraction pattern measured rather than the sample absorption; this enabled the identification of the crystallographic orientation for all parts of the sample. First, our DCT measurements indeed reveal a single orientation throughout the entire sample, confirming the single crystallinity of our samples. Next, pole figures were generated for $<\!0001\!>$ (Supplementary Fig.~\ref{fig:DCT}c), $<\!10\bar10\!>$ (Supplementary Fig.~\ref{fig:DCT}d), and $<\!2\bar1\bar10\!>$ (Supplementary Fig.~\ref{fig:DCT}e) of the single-crystalline sample, indicating the directions of the crystallographic vectors in relation to the Cartesian coordinate system of the 3D model. These vectors were then visually matched to the physical sample (Supplementary Fig.~\ref{fig:DCT}a). The sample was then cut and polished into rectangular prisms with the length along the desired crystallographic direction.

\clearpage
\section{Supplementary Note 8: Additional Samples}
To explore directional dependence and variation between samples, we performed AHE measurements on 4 separate samples, representing at least 2 different orientations within the \textit{ab}-plane. Table~\ref{tab:sample_table} lists the samples measured along with their orientations and dimensions while Table~\ref{tab:condition_table} lists and describes the mounting conditions. 

\begin{table}[h]
    \centering
    \begin{tabular}{ |c|c|c|c|c| } 
    \hline
    Sample & Orientation & Length ($\mu$m) & Width ($\mu$m) & Thickness ($\mu$m) \\ 
    \hline
    A & $[10\bar10]$ & 1338 & 542 & 194 \\
    \hline
    B & $[10\bar10]$ & 1301 & 488 & 201 \\
    \hline
    C & 9\degree from $[10\bar10]$ & 1290 & 538 & 200 \\
    \hline
    D & Unknown & 1237 & 514 & 187 \\
    \hline
    \end{tabular}
    \caption{\textbf{List of samples and their orientations.}}
    \label{tab:sample_table}
\end{table}

\begin{table}[h]
    \centering
    \begin{tabular}{ |c|c| } 
    \hline
    Condition & Description \\ 
    \hline
    Ambient & Mounted on ETO PPMS puck with GE varnish \\
    \hline
    Mounted & Mounted over gap of stress cell, 0.00\% mechanical strain \\
    \hline
    Compressed & Mounted over gap of stress cell, -0.118\% mechanical strain \\
    \hline
    \end{tabular}
    \caption{\textbf{List of sample conditions and their descriptions}}
    \label{tab:condition_table}
\end{table}

\begin{figure} [ht]
    \includegraphics[width = 6in]{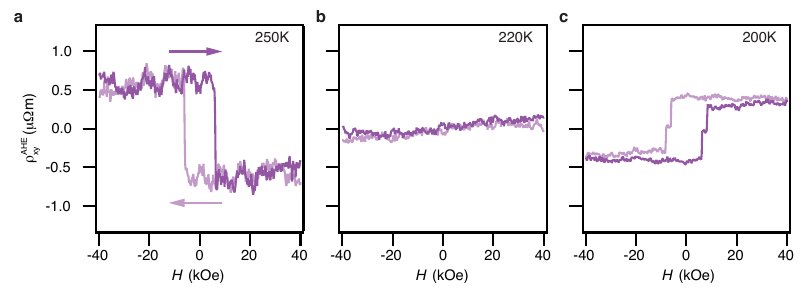}%
    \caption{\textbf{Switching of the AHE in Sample C.} 
    \textbf{a}-\textbf{c}, Anomalous Hall resistivity as a function of magnetic field at 250\,K (\textbf{a}), 220\,K (\textbf{b}), and 200\,K (\textbf{c}). The arrows in (\textbf{a}) represent the direction of field sweep during the measurement.
    \label{fig:SampleC}}
\end{figure}

Supplementary Figure~\ref{fig:SampleC} shows the AHE at three different temperatures for Sample C in the Mounted condition. For this sample, the current/strain was directed roughly $9\degree$ from $[10\bar10]$. We again observe a clear switch in the sign of the AHE with a critical temperature of ~225\,K, in agreement with the other samples measured.

\begin{figure} [ht]
    \includegraphics[width = 6in]{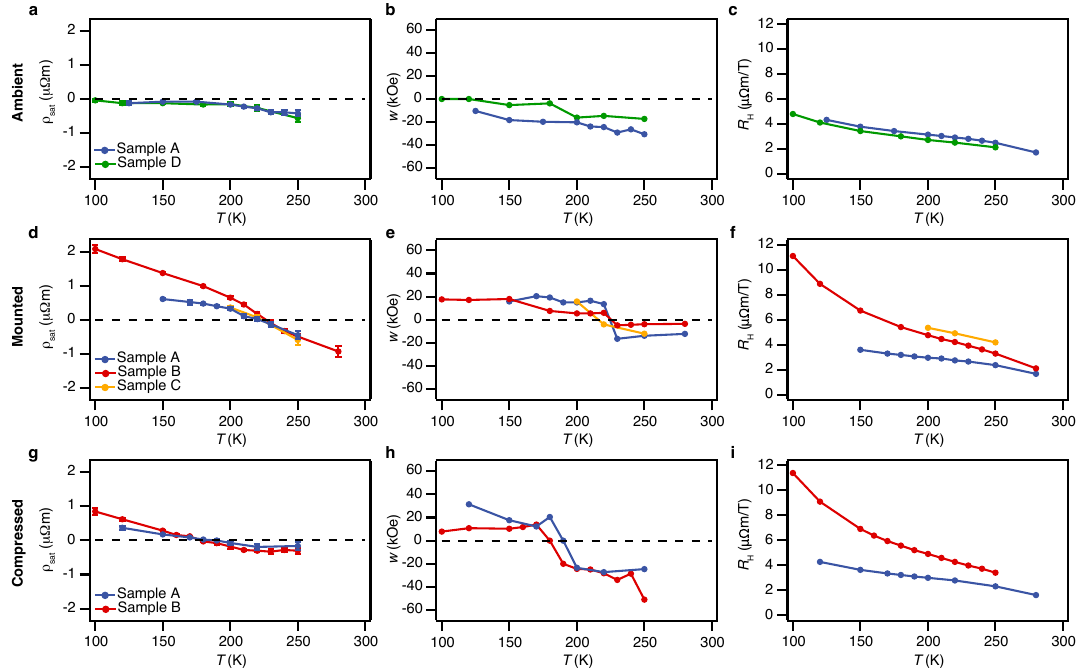}%
    \caption{\textbf{Comprehensive comparison of Hall behavior across samples, strains, and temperatures.} 
    \textbf{a}-\textbf{c}, Saturated Anomalous Hall resistivity $\rho \textsubscript{sat}$ (\textbf{a}), AHE width $w$ (\textbf{b}), and Hall coefficient $R_{\text{H}}$ (\textbf{c}) as a function of temperature for all samples measured in the Ambient condition. \textbf{d}-\textbf{f}, Same as (\textbf{a})-(\textbf{c}) but for the Mounted condition. \textbf{g}-\textbf{i}, Same as (\textbf{a})-(\textbf{c}) but for the Compressed condition.
    \label{fig:All_Sample_Comp}}
\end{figure}

We present a comprehensive comparison of the saturated anomalous Hall resistivity in the $+H$ polarized state $\rho_\text{sat}$, width $w$ (width of hysteresis loop at $\rho_{\text{sat}}=0$), and Hall coefficient $R_\text{H}$ at the three separate strain conditions for all samples measured (Supplementary Fig.~\ref{fig:All_Sample_Comp}). While we observe slight variations from sample to sample, the overall behavior and magnitudes of $\rho_\text{sat}$ and $w$ are similar for all samples. The slightly larger magnitude of $\rho_\text{sat}$ in Sample B can be explained by the higher $R_\text{H}$, indicating a lower carrier concentration. Despite slight changes in $R_\text{H}$, the overall trends and transition temperatures remain from sample to sample. More specifically, the transition temperature of the AHE sign flip $T^*$ all occur within $<3$\,K of 224\,K for the Mounted condition and between 180 and 190\,K for the compressed condition. Notably, no significant differences in magnitude nor transition temperature are observed between Sample A ($[10\bar10]$ orientation) and Sample D (random orientation within the \textit{ab}-plane) in the Ambient condition or Sample A and B ($[10\bar10]$ orientation) and Sample C (9\degree from $[10\bar10$ orientation) in the Mounted condition. This indicates that the direction of applied stress in the \textit{xy}-plane has minor impact on the AHE switching behavior observed in MnTe, suggesting that $\varepsilon_{zz}$ dominates the strain dependence of the AHE.

\clearpage

\clearpage
\section{References}

\normalem
%

\end{document}